# On the Fundamental Limits of Single-snapshot Compressive Ultrasound Imaging Using Random Aberrative Masks

Zehua Dou *, Yaokuan Zhang, Jialong Zhang, Cherif Othmani, Lars Büttner, and Jürgen W. Czarske

***Abstract*—Compressive sensing emerges as a paradigm shift to realize real-time (> 100 Hz) volumetric ultrasound imaging using only a single element transducer equipped with an aberrative mask. Such a coded aperture converts each scatterer within the field of view as a specific echo signal, effectively compressing the volumetric information into time sequences that can subsequently be reconstructed into images via computational methods. Thus, compressive imaging can greatly reduce the data rate and simplify the electronics. Despite this potential, the practical performance of single-snapshot compressive ultrasound imaging remains insufficiently understood, particularly how this is influenced by coded aperture design, imaging task complexity, and reconstruction strategy. To address this gap, we separate the information budget provided by the aperture from the algorithm-dependent extraction of the encoded information. First, the spatial impulse responses of random aberrative masks with different pixel sizes and time-delay ranges were experimentally calibrated, and the entropy-based effective rank of their similarity matrix was used to quantify the available encoding capacity. Masks with pixel size of approx. half wavelength and a time-delay range of two carrier periods provided higher encoding capacities, reaching up to 1.1% of the total sampled voxels. Algorithm-dependent information extraction was further evaluated for both ULM-motivated particle localization and B-mode imaging. For particle localization, L1-norm regularized least-squares method faithfully reconstructed particles corresponding to up to approx. 10% of the available encoding capacity, substantially outperforming the matched filter. A transition from successful to failed reconstruction when the number of particles exceeded this value was observed, revealing the upper limit of the present compressive imaging systems using random masks. For B-mode imaging, LSQR achieved the highest structural similarity index measure among the evaluated methods, of up to 0.12, although the encoding capacity of the current random masks remained insufficient for high-fidelity reconstruction. This work provides a quantitative basis and a unified framework to interpret the recoverability of compressive imaging systems, guiding future mask optimizations toward greater encoding orthogonality and more efficient information extraction.**

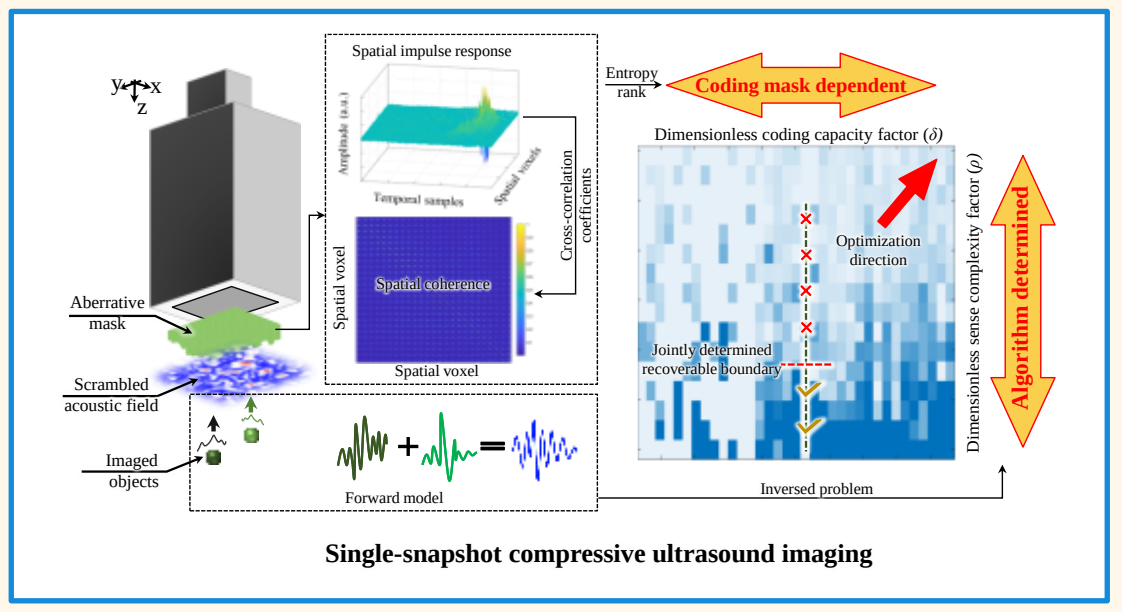

## I. Introduction

FOUR-DIMENSIONAL (4D) ultrasound imaging in large volumes e.g., over 10 $cm^3$ is greatly envisioned for a wide range of biomedical imaging [1], [2], healthcare monitoring [3], human machine interface [4], [5], and technical processes monitoring applications [6]–[11]. Conventional volumetric (3D) ultrasound imaging systems rely on either mechanical scanning of ultrasonic transducers [12]–[14], or (micro-) beamforming using matrix transducer arrays [15]–[17]. In volumetric imaging applications, the temporal resolution of the former is limited by the mechanical scanning process in large field of view (FoV) [13]. While the imaging speed of the latter is constrained by the transferring and processing the heavy data e.g., above 20 GB/s, generated by phased arrays with over 1000 channels for covering the desired FoV [18]. In short, conventional ultrasound imaging techniques are inherently compromised by the trade-off between imaging performance and FoV dimension, meanwhile relying on bulky and costly

Pre-print submitted for journal peer review. This work was supported by the German Aerospace Center Project Sponsor (DLR-PT) within the support program "Industrial JointResearch" (IGF) of the German Federal Ministry of Economics and Technology under Grant 1F23280N, and in part by German Research Foundation (Deutsche Forschungsgemeinschaft DFG) under Grants 512483557 (BU2241/9-1) and 411799900 (CZ5543-02). (* *Corresponding author: Zehua Dou*)

Zehua Dou, Yaokuan Zhang, Jialong Zhang, Lars Büttner, Cherif Othmani and Jürgen W. Czarske are with the Faculty of Electrical and Computer Engineering, Laboratory of Measurement and Sensor System Technique, Technische Universität Dresden, 01069 Dresden, Germany (e-mails: zehua.dou@tu-dresden.de, z.dou@ifw-dresden.de; yaokuan.zhang@mailbox.tu-dresden.de; jialong.zhang@mailbox.tu-dresden.de; lars.buettner@tu-dresden.de; cherif.othmani@tu-dresden.de; juergen.czarske@tu-dresden.de).

***Highlights***

- **A framework linking coding mask design and reconstruction scheme with single-snapshot compressive ultrasound imaging performance.**
- **A pipeline for quantifying the space bandwidth product that governs imaging performance in practical applications.**
- **A systematic evaluation of single-snapshot imaging performance using random aberration masks towards ultrasound localization microscopy and B-mode imaging scenarios.**

electronics.

To overcome the above limitations, novel imaging methods using reduced number of transducers are greatly demanded. Recent developments of sparse arrays [19], row-column addressed arrays [20] and lensed arrays have demonstrated promising strategies for reducing the active channel counts by one to two orders of magnitude [21]. These methods mostly rely on the time-of-flight model to coherently align echoes or transmitted signals from different channels, and they therefore require dense elements for spatial sampling and often redundant transmission and receiving events to achieve robust focusing.

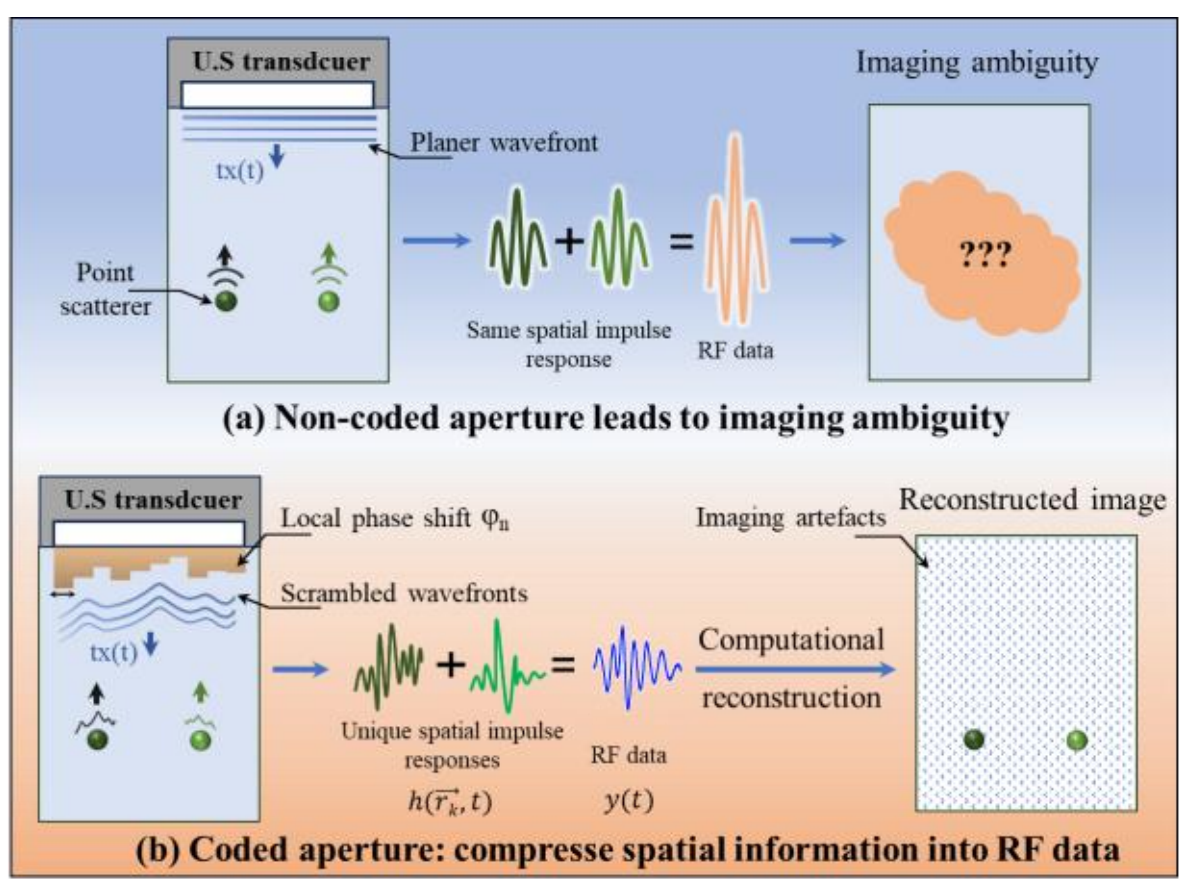

Fig. 1. Schematic diagram of compressive ultrasound imaging using a single element transducer through aberrative mask. (a) A non-coded aperture results to imaging ambiguity due to an incapability of distinguishing objects from specific radio-frequency (RF) signals. (b) A coded aperture is able to compress spatial information into a RF signal by introducing a set of unique spatial impulse responses.

In contrast, compressive sensing marks a striking paradigm shift toward computational imaging, solving the inversed problem of wavefields, which enables volumetric image reconstruction from highly limited measurements e.g., using only a single element transducer [22], [23]. In short, a compressive imaging system employs an aberrative mask that introduces local time delay, effectively producing phase modulation of coherent signals. This allows for transmitting and receiving spatiotemporally scrambled ultrasonic field that can encode each scatterer as a specific echo signal i.e., spatial impulse response function, thereby compressing each volume as a time sequence. The images are then reconstructed using computational methods (see **Fig. 1**) e.g., matched filter/time-reversal scheme [24]. Thus, single-snapshot compressive imaging enables fast volumetric acquisitions that are primarily defined by the transient time of double-path wave propagation, meanwhile using single channel excitation and data acquisition electronics, which is greatly simplified compared with conventional systems.

However, existing studies predominantly adopt a system specific perspective, where the imaging performance is often assessed only for specific coding mask designs or reconstruction algorithms [25]–[27]. As a result, the obtained insights may be difficult to generalize across other implementations of compressive imaging systems. To sum up, there is currently a lack of fundamental understanding of the limit of single-snapshot compressive imaging, which is critical towards guiding system design and parameter selection in practical applications e.g., concentration of tracer particles in ultrasound localization microscopy (ULM).

To address this gap, in this work, we experimentally characterize the spatial impulse response functions of random aberrative masks with different designs, in terms of pixel sizes and time delay ranges, under single-snapshot pulse-echo mode. The encoding capability of a mask is determined by the mutual coherence within its spatial impulse response functions. Building upon this foundation, the reconstruction performance of relevant methods is systematically evaluated, including matched filter as the baseline, sparsity-constrained least-squares (LS) reconstruction particle tracking towards ULM [28], as well as LS and total variation (TV) regularized LS reconstructions for B-mode imaging [29]. As a result, the spatial encoding and reconstruction capabilities jointly determine the number of voxels that can be simultaneously resolved in practical scenarios, which defines an effective space bandwidth product (SBP) of a compressive imaging system. As the effective SBP quantifies the practical upper limit of a system, it manifests itself as a transition between successful and failed reconstruction regimes when the scene complexity exceeds this limit.

Given these, this work contributes to a unified framework linking the coded aperture design and reconstruction algorithm with single-snapshot compressive ultrasound imaging performance; a pipeline for quantifying the SBP of a compressive imaging system that governs the performance in practical imaging tasks; and a systematic evaluation of imaging performance using random aberrative masks towards both ULM and B-mode (structural) imaging applications.

Following the introduction, the rest of this paper is organized in four sections: **Section II** mathematically formulates the problem of compressive imaging, including the spatial information encoding process, the information capacity analysis of coding masks and the reconstruction methods addressed in this work; **Section III** describes the different random coding masks and the experiment procedures for acquiring the spatial impulse response functions of the different masks; **Section IV** systematically evaluates the information capacities of these different coding masks and the joint performance of reconstruction schemes for both ULM and B-mode imaging using digital phantoms; **Section V** concludes the paper by summarizing the fundamental limits of single-snapshot compressive imaging using random aberrative masks and discusses insights into optimal coded aperture design.

## II. Mathematical Formulation

The basic principle of compressive imaging using a single-element transducer is illustrated in **Fig. 1**. Successful implementation of such a system relies on two key aspects, namely appropriate design of the aberrative mask for efficient encoding of spatial information to ensure sufficient information capacity under limited measurements, and high-fidelity reconstruction that preserves the encoded information. This section provides the mathematical formulation of these two aspects.

### *A. Aberration mask Spatial information encoding and information capacity of coding masks*

The essence of encoding spatial information is to generate a set of unique spatial impulse responses of individual point scatterers within the measured volume, such that the scatterers located at different positions can be afterwards distinguished unambiguously. For this, an aberrative mask must be employed, on which a specific time delay is applied locally to each pixel, for distorting the axially symmetric planar wavefronts of a single element transducer. Thanks to these unique spatiotemporal signatures, an entire imaging volume can be effectively encoded and compressed into one radio-frequency (RF) signal. This encoding and compression process can therefore be mathematically written in (1), in which $y(t)$ is a column vector of RF data, $\boldsymbol{x}$ is the measured volume arranged into a column vector, $\boldsymbol{A}$ is a matrix containing each spatial impulse response function $a(\vec{r},t)$ at corresponding column, and $n(t)$ denotes the additive noise [24].

$$y(t) = \underbrace{[a(\vec{r}_1,t), a(\vec{r}_2,t), \cdots, a(\vec{r}_n,t)]}_{\boldsymbol{A}} \cdot \underbrace{\begin{bmatrix} x_1 \\ x_2 \\ \vdots \\ x_n \end{bmatrix}}_{\boldsymbol{x}} + n(t) \quad (1)$$

Such encoding of spatial information is able to drastically reduce the number of excitation and acquisition events and simplify the electronics, thereby eliminating the data transferring and(or) mechanical scanning bottlenecks in conventional imaging systems. Moreover, the volumetric acquisition rate is determined by the duration of the RF signal that needs to be recorded, which is approximately the transient time of double-path wave propagation. For instance, the volume acquisition time of single-snapshot compressive imaging at pulse-echo mode can be as fast as approx. 134 µs/volume (i.e., above 7.4 kHz) for an imaging depth of 10 cm in water.

To determine the designs of aberrative masks that can enable efficient encoding of spatial information, the information capacities provided by a coding mask and that required in ultrasound imaging must be analysed. In pulse-echo mode ultrasound imaging, the lateral resolution is determined by the ultrasonic wavefront profile through a mask, while the axial resolution is only defined by the pulse length, independent of the mask [13]. Thus, the following analysis is conducted in the lateral direction only.

On one hand, the information capacity required by an imaging system can be quantified by the desired SBP, which is calculated by the product of imaging volume and spatial bandwidth of an image. In 2D lateral space, this relation is given in (2), in which $L_{img}^2$ denotes the area of image, and $B^2$ is the spatial bandwidth of the image (unit of $1/\mathrm{m}^2$).

$$SBP_{image} = L_{img}^2 \cdot B^2 \quad (2)$$

Since the spatial bandwidth of an image is the reciprocal of its spatial resolution, the upper bound of an image's spatial bandwidth is obtained from the diffraction limit at the employed ultrasonic frequency, as given in (3) [13]. Thus, the SBP of an imaging system indicates the number of resolvable pixels (or voxels) in the FoV.

$$B \leq 2/\lambda \quad (3)$$

On the other hand, the number of linearly independent vectors in $\boldsymbol{A}$ i.e., spatial encoding modes, provides an SBP-like measure of the information capacity of a coded aperture. However, the exact value of this metric depends on the complete wavefield formation process, which is influenced by multiple factors e.g., mask design, bandwidth, propagation distance, to name a few. It is therefore difficult to directly predict the diffraction based on a simple mask design parameter. As a first-order geometrical approximation, the information capacity of a random coding mask is estimated here by the number of independently controllable pixels on a mask [30], as given in (4) where $L_{mask}^2$ and $d^2$ are the areas of the mask and of the pixel (unit of $\mathrm{m}^2$), respectively. A more rigorous derivation and characterization of the actual information capacity of a mask is presented in **section IV**.

$$SBP_{mask} = L_{mask}^2 / d^2 \quad (4)$$

Aiming for efficient spatial information encoding, the pixel size of a coding mask must be designed accordingly, to match with the desired SBP of a compressive imaging system. For instance, assuming that the mask and image areas are identical i.e., $L_{img}^2 = L_{mask}^2$, to realize a lateral resolution at the diffraction limit of ~ $\lambda/2$, the pixel size on a mask must obey (5) [31].

$$SBP_{mask} \geq SBP_{image} \xrightarrow{yields} d \leq \lambda/2 \ @ \ diffraction\ limit \quad (5)$$

Note that the inequity in (5) considers the diffraction in reality i.e., the wave field slightly spreads in the lateral direction. As a result, the effective FoV of a mask is slightly greater than its geometrical dimension.

To unveil the impact of mask design on the spatial encoding performance i.e., the information capacity, we investigated random coding masks with different pixel sizes and time delay ranges. The details of mask designs are given in **Section III**.

### B. Inverse problem

With prior knowledge of the spatial impulse response functions introduced by a coded aperture i.e., matrix $\boldsymbol{A}$, the images can then be restored from the measured RF signal, using computational algorithms, solving the inverse problem of (1). Because single-snapshot compressive imaging relies on highly limited measurements, the inverse problem is generally ill-conditioned, and the reconstruction quality therefore depends not only on the encoding capability of the mask, but also on the reconstruction algorithms. Moreover, different methods solving the inverse problem are suitable for different imaging scenarios.

#### 1) Matched filter

The matched filter provides a computationally efficient method for identifying known signals with the highest signal-to-noise ratio (SNR) against additive white Gaussian noise [32]. The matched filter projects a measured RF signal back to the spatial domain via the $A$ matrix, as given in (6), in which $A^H$ and $A^H{\cdot}A$ are the conjugated transpose and the Gram matrix of $A$ matrix, respectively.

$$\boldsymbol{x_{MF}} = \boldsymbol{A^H} \cdot y(t) = \underbrace{\boldsymbol{A^H} \cdot \boldsymbol{A}}_{Gram} \cdot \boldsymbol{x} + \underbrace{\boldsymbol{A^H} \cdot n(t)}_{noise} \quad (6)$$

Moreover, the Gram matrix is a square matrix, in which each diagonal and off-diagonal elements represent the auto-correlation and cross-correlation of the corresponding spatial impulse response(s), respectively. In other words, the Gram matrix reflects the orthogonality among the spatial impulse response generated by a coded aperture. Thanks to the computational efficiency of the Gram operation, it serves as a baseline for assessing the impact of encoding capability on the imaging performance.

#### 2) Sparsity-constrained least-squares reconstruction for particle localization towards ULM applications

In ultrasound localization microscopy (ULM), each integrated frame consists of a small number of isolated tracer particles, which can be considered as sparse in the spatial domain. With such a prior assumption, the reconstruction can therefore be formulated as an sparsity-constrained least-squares (LS) problem, as given in (7).

$$\boldsymbol{x_{L1}} = \underset{x}{argmin} \frac{1}{2} \|\boldsymbol{A} \cdot \boldsymbol{x} - y(t)\|_2^2 + \lambda \|\boldsymbol{x}\|_1 \quad (7)$$

The first term in (7) ensure the consistency between the measured RF signal and the one corresponding to the estimated image ($x$). The second term promotes sparsity in the reconstructed image, in which $\lambda$ is a weighting factor that controls the contribution of the L1-norm regularization. In this work, the L1-regularized LS problem is solved using the fast iterative shrinkage thresholding algorithm (FISTA) that is initialized using the matched filter result, with $\lambda$ value of 0.05 and 500 iterations. The iteration process is briefly summarized as follow. In each iteration $k$, FISTA first calculates an intermediate gradient-descent step $v_k$ based on the accelerated variable $z_k$, as given in (8), where $L$ is the Lipschitz constant of the gradient of the LS term [33].

$$\begin{cases} v_k = z_k - (A^H \cdot (A \cdot z_k - y))/L \\ L \geq \|A^H \cdot A\|_2 \end{cases} \quad (8)$$

The sparsity of the updated image estimation $x_{k+1}$ is then forced by applying an element-wise soft-thresholding i.e., the so-called shrinkage operator, as given in (9).

$$x_{k+1} = sgn(v_k) \cdot max(|v_k| - \lambda/L, 0) = \begin{cases} 0, |v_k| \leq \lambda/L \\ v_k - sgn(v_k) \cdot (\lambda/L), |v_k| > \lambda/L \end{cases} \quad (9)$$

Finally, the accelerated variable $z_{k+1}$ for the next iteration is updated according to (10), which is known as a momentum update step.

$$\begin{cases} z_{k+1} = x_{k+1} + \beta_k (x_{k+1} - x_k) \\ \beta_k = (t_k - 1)/t_{k+1} \\ t_{k+1} = \left(1 + \sqrt{1 + 4t_k^2}\right)/2 \end{cases} \quad (10)$$

The performance of reconstructing sparse images under a given encoding capacity is crucial to determine the tracer concentrations and single frame reconstruction time, which jointly determine the required integration time in practical ULM imaging applications.

#### 3) Least-squares reconstruction

In B-mdoe (structural) imaging, the objects are generally not sparse, and the sparsity regularization may not be appropriate. Therefore, one way of structural imaging is to estimate an image with the minimum residual between its forward projection in time domain and the measured RF signal. This can be formulated as a pure LS problem, as given in (11).

$$\boldsymbol{x_{LS}} = \underset{x}{argmin} \frac{1}{2} \|\boldsymbol{A} \cdot \boldsymbol{x} - y(t)\|_2^2 \quad (11)$$

In this work, this LS problem is solved using the LSQR algorithm [34], with the default realization in MATLAB R2024b [35]. Physically, the LSQR algorithm can be interpreted as a global deconvolution process with respect to the Gram matrix as the kernel that encodes the inter-voxel crosstalk.

#### 4) Total variation (TV) regularization for B-mode imaging with clear boundaries

In addition to pure LS reconstruction, B-mode imaging may also be benefited from reasonable spatial regularizations e.g., for noise reduction and blurring suppression [36]. In particular, the objects in B-mdoe imaging e.g., inclusions, channels and interfaces, may possess connected regions with well-defined boundaries. Total variation (TV) regularization is particular suitable for these cases where large-scale boundaries should be preserved, while small-scale speckle patterns and reconstruction artifacts are suppressed. As its name suggests, TV constrains the spatial variations in the images. In 2D images, the TV regularized problem is written as (12) [36].

$$\begin{cases} \boldsymbol{x_{TV}} = \underset{x}{argmin}\,\frac{1}{2}\|\boldsymbol{A}\cdot\boldsymbol{x} - y(t)\|_2^2 + \lambda\,\underbrace{\|\nabla\boldsymbol{x}\|_1}_{TV(x)} \\ TV(\boldsymbol{x}) = \sum_{i,j}\sqrt{\left(x_{i+1,j} - x_{i,j}\right)^2 + \left(x_{i,j+1} - x_{i,j}\right)^2} \end{cases} \quad (12)$$

In this work, TV regularization is realized by the FISTA algorithm as well, with $\lambda$ value of $1 \times 10^{-4}$ and 10 iterations. Specifically, the gradient-descent and the momentum update steps remain the same as given in (8) & (10). While, the shrinkage operator is replaced by the TV proximal operation to suppresses local spatial variations while preserving large-scale boundaries, as given in (13) that is solved via the Chambolle's dual projection algorithm [37].

$$x_{k+1} = \underset{x}{argmin}\,\frac{1}{2}\|\boldsymbol{x} - v_k\|_2^2 + (\lambda/L)\cdot TV(x) \quad (13)$$

The reconstruction performance of these methods towards ULM and B-mode imaging are thoroughly investigated in **Section IV**. The matched filter serves as a baseline for both application scenarios. The sparsity-constrained LS problem is dedicated for particle localization towards ultrasound localization microscopy, while the LSQR and TV-regularized LS problems are evaluated for B-mode imaging. Together with the encoding capability of the aberrative masks, the effective SBP of a single-snapshot compressive ultrasound imaging system can be determined. In this work, reconstructions were performed slice-by-slice at the calibrated imaging depths, as in most existing compressive imaging studies. This is because a full volumetric forward matrix containing the spatial impulse responses of all voxels in the volume leads to prohibitively large memory and computational requirements.

## III. Experimental Setup and Procedure

This section presents the coding masks used in the experiments and the experimental procedure for characterizing the spatial impulse responses of each mask in the FoV.

### A. Coding mask design and fabrication

In order to elucidate the influences of coding mask designs and encoding capability, we designed and fabricated in total 5 different masks with pixel sizes (side length) of 100 µm, 150 µm and 300 µm, as well as time delay ranges of 0.2 µs and 0.4 µs, as listed in **Table I** and shown in **Fig. 2**. Given the small pixel sizes, we used high resolution digital light processing (DLP) technology to fabricate these coding masks (Boston Micro Fabrication S240 DLP-machine). This machine provides high printing resolution of 10 µm and 30 µm in lateral and axial directions, respectively. These masks possess a same diameter of 10 mm, with squared pillars with different heights as pixels. Thanks to the different speed of sound in the mask material and the imaging medium e.g., water, each pixel induces a local time delay to the transmission and receiving paths.

Table I. The Configurations of The Different Coding masks

| Mask number | Pixel size (µm) [a] | Time delay range [b] |
|---|---|---|
| #1 | 100 (~ $\lambda$/3) | 0.2 µs ($\tau$) |
| #2 | 100 (~ $\lambda$/3) | 0.4 µs ($2\tau$) |
| #3 | 150 (~ $\lambda$/2) | 0.2 µs ($\tau$) |
| #4 | 150 (~ $\lambda$/2) | 0.4 µs ($2\tau$) |
| #5 | 300 (~ $\lambda$) | 0.4 µs ($2\tau$) |

[a] The wavelength $\lambda$ corresponds to the central frequency of 5 MHz.
[b] The time delay range refers to the value in water, and with respect to the period ($\tau$) of the central frequency of 5 MHz.

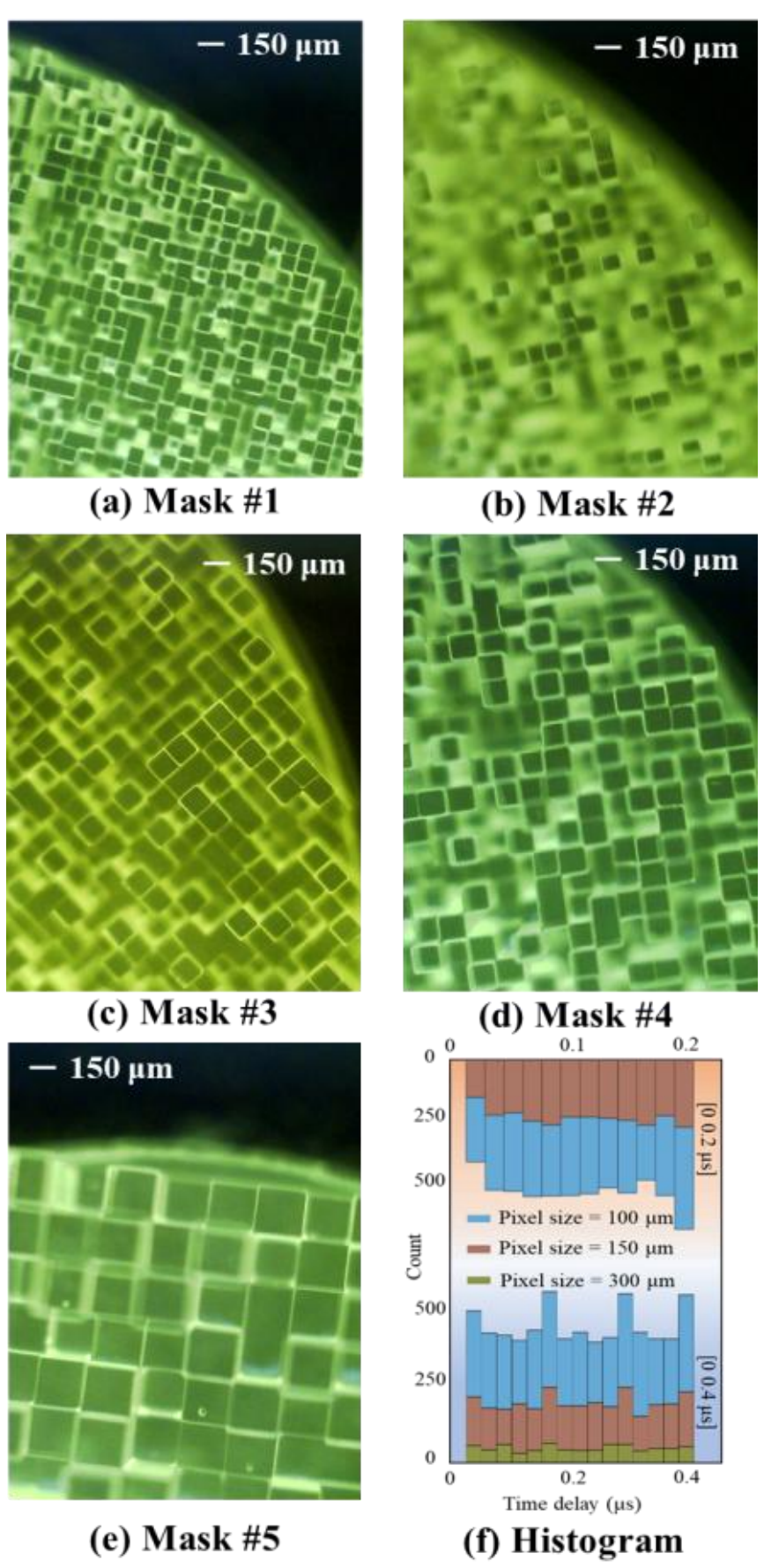

Fig. 2. The five random aberrative masks used in the experiments and their histograms.

To design the time delays of pixels on the mask, we first characterized the acoustic properties of the cured DLP resin using pulse echo measurements with bulk samples. As a result, the density was 1.289 g/cm$^3$, the longitudinal wave velocity was 2557.54 m/s, and the attenuation coefficient was 2.72 dB/(MHz·cm). Based on the value of longitudinal wave velocity mismatch in SLA resin and in water, the required heights of pillars as pixels for enabling time delays in ranges of 0 μs - 0.2 μs and 0 μs - 0.4 μs were calculated as approx. 0.77 mm and 1.53 mm, respectively. We assigned random heights, following independently and identically distributed (i.i.d) uniform random numbers in these two ranges, to each pillar on the masks. To be printable, these random numbers were rounded to integer multiples of the given axial resolution i.e., 30 μm. **Fig. 2** (d) shows the histograms of the time delays on the 5 masks, revealing their nearly i.i.d uniform distributions.

### B. Signal excitation and acquisition

To investigate the imaging performance of the 5 masks, we first calibrated their spatial impulse responses in deionized (DI) water. For this, we mounted the different masks on a single element ultrasonic transducer with a centre frequency of 5 MHz with a fractional bandwidth of 60 %, and an active diameter of 6 mm (Sonotec GmbH, Germany), as shown in **Fig. 3**. A needle hydrophone with a diameter of 200 μm (Precision Acoustics, UK) was used as a point reflector. The hydrophone was mounted on an in-house-developed translational scanning system, and we performed raster scans over 12 mm × 12 mm planes that are coaxially aligned with the coded aperture. The echo signals were recorded at all the scanned positions that are spaced by a distance of 150 μm i.e., in total 6561 (81 × 81) acquisitions from a scanning plane. For assessing the volumetric imaging performance, the spatial impulse responses of each mask were recorded on 7 equidistant planes, at distances of 1 mm to 31 mm from the mask surface, spaced by 5 mm apart. This distance region covers the near field and transition region of the single element transducer.

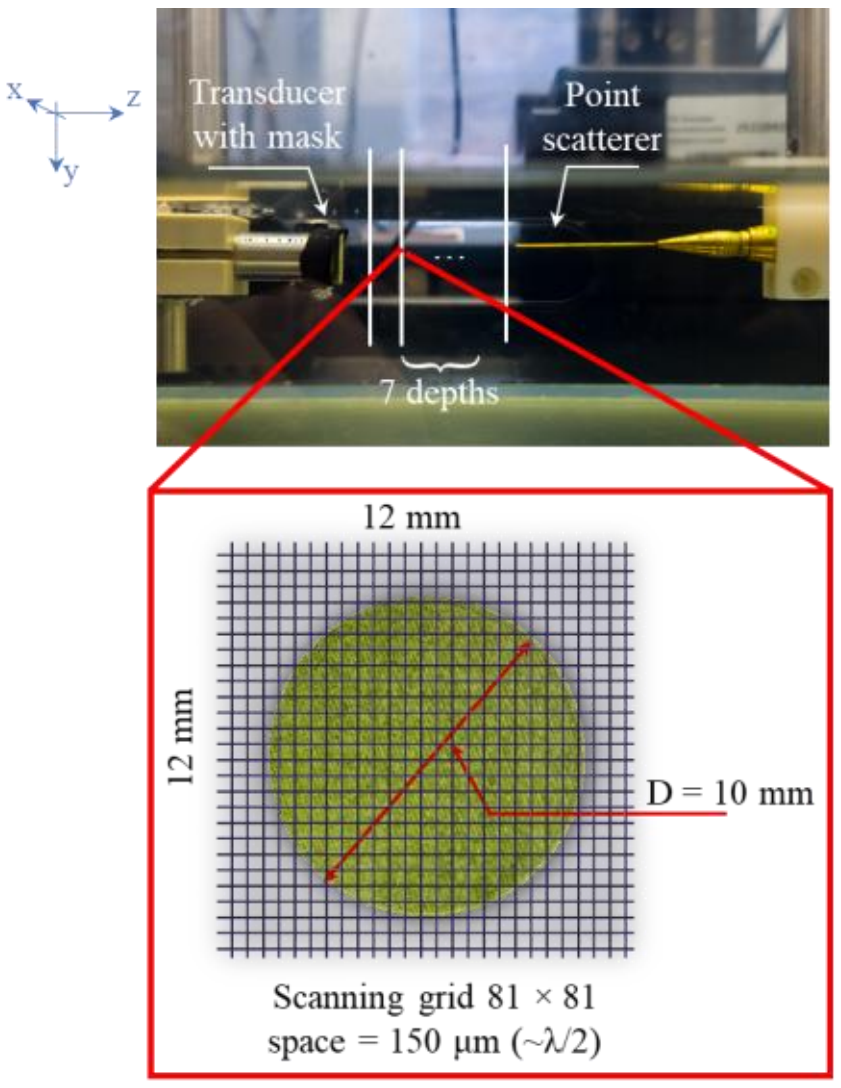

Fig. 3. Experimental setup and procedure for calibrating the spatial impulse response functions of the different masks, in which the coordinates used throughout this paper is defined.

The coded aperture was excited by 3-period 5 MHz tone bursts at a pulse repetition frequency of 2 kHz, using a US4R-lite research platform. The echo signals were recorded for 4096 samples at a sampling frequency of 65 MHz, ensuring full coverage of the scanning region. Due to the relatively high attenuation coefficient of the 3D printing material, a relatively high voltage of 50 V was used, and 20 acquisitions at each scanned position were averaged for increasing the SNR. After data averaging, the signals were bandpass filtered within 3 – 6 MHz. These treatments ensure a final SNR of above 20 dB, which is the basis of accurately extracting the spatial impulse responses.

## IV. Results and Discussions

This section starts with the characterizations of the acoustic fields and the spatial encoding capabilities of the five masks, introducing an SBP-like measure of their information capacities. On this basis, the performance of single-snapshot compressive imaging using the different reconstruction methods are further investigated. Notably, we introduce two dimensionless parameters, namely the dimensionless coding capacity factor $\delta$ and the dimensionless scene complexity factor $\rho$, to describe the system-side information budget and the scene-side reconstruction load, respectively. These parameters allow the imaging performance of different masks, reconstruction methods, and imaging tasks to be evaluated under a unified framework.

### A. Wave field through coding mask

As a first demonstration of the spatial information encoding effect enabled by the masks, **Fig. 4** compares the acoustic wavefronts generated by non-coded and coded apertures. The wavefronts shown here are the instantaneous amplitudes of the transmitted signals recorded by the hydrophone on a lateral scanning plane. It is observed that without the coding mask, the single element transducer only produced a nearly planar wavefront that is mostly symmetric about the axial axis, leading to imaging ambiguity. In contrast, the coded aperture distorts the transmitted wavefront and breaks the axial symmetry, which indicates the feasibility of spatial information encoding.

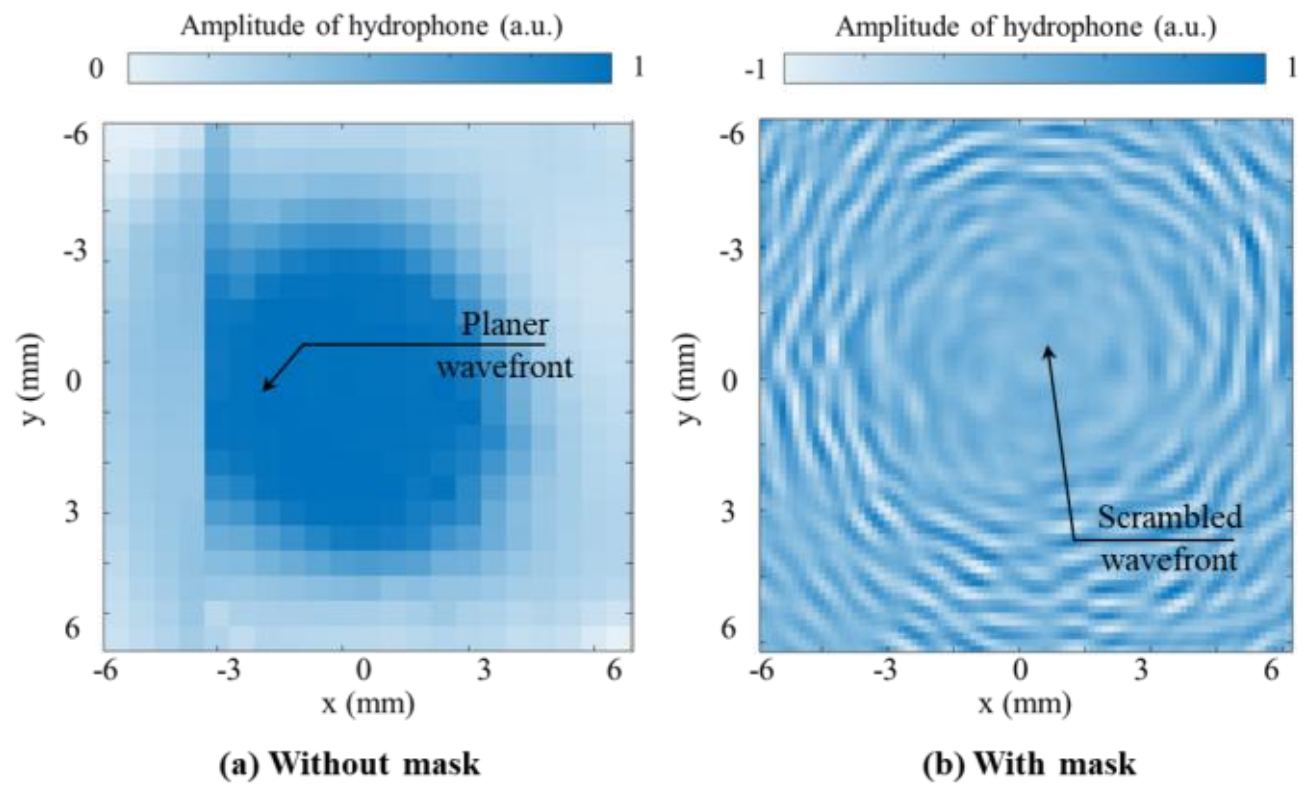

Fig. 4. A comparison of the wavefronts generated by a single transducer without and with a coding mask (mask #3 at 11 mm).

**Fig. 5** further compares the spatial distributions of the echo energy of the different masks, using the -10 dB contours of their pulse-echo response energy. The comparison shows that the spatial distribution of response energy is strongly affected by the mask design. For the two masks with the smaller time-delay range of 0.2 μs i.e., one period of the carrier frequency, the high energy region remains relatively concentrated around the axial axis of the aperture, whereas the masks with the larger time-delay range of 0.4 μs produce a broader and more fragmented sensitivity distribution. Although the amplitudes of the response in the regions outside this contour are much lower, they still contribute to the measured RF signal. Such amplitude variations assign different weights to different voxels and may influence forward model conditioning, and therefore the reconstruction robustness. This behaviour is attributed to mask-induced spatial constructive and destructive interference of the transmitted and received wavefields. However, the detailed interference mechanism depends on the specific random realization of the mask, and is beyond the main focus of this current work.

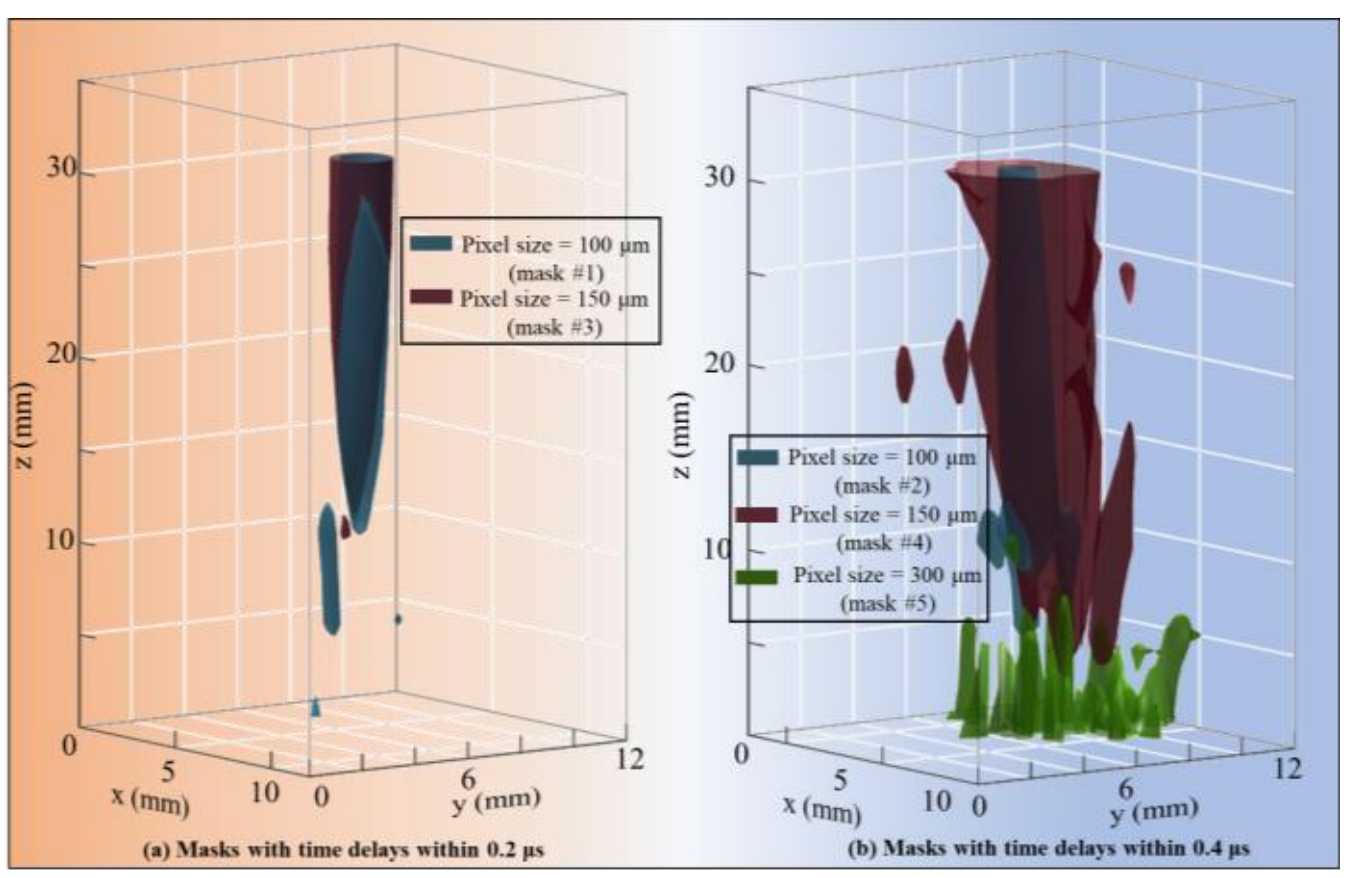

Fig. 5. The sensitivity responses of the different coding masks characterized by the -10 dB contours of the pulse-echo response energy: (a) the masks with time-delay range of 0.2 μs, and (b) the masks with time-delay range of 0.4 μs.

## B. Encoding capability

### 1) Spatial coherence length among spatial impulse response

As a first impression of the ability to encode the different voxels as unique RF signals, **Fig. 6** (a) plots a set of exemplary spatial impulse responses generated by a coded aperture at a given depth i.e., the $z$ coordinate. This set of responses cover 6561 spatial points, while their effective RF responses occupy mainly a temporal window of approx. 200 samples. As a result, the spatial information must be encoded into a limited number of temporal samples that is much less than the number of independent voxels. In other words, these 6561 spatial impulse responses cannot be fully orthogonal with each other, unavoidably leading to inter-voxel coherence. Such spatial coherence among the spatial impulse response matrix $\boldsymbol{A}$ is the origin leading to the ill-posed inverse problems for image reconstruction.

To evaluate the spatial coherence, we introduce the cosine similarity matrix of the spatial impulse responses. Each element in this matrix $\boldsymbol{C_{i,j}}$ represents the pairwise correlation coefficient, as given in (14).

$$\boldsymbol{C_{i,j}} = \frac{\boldsymbol{a_i^H} \cdot \boldsymbol{a_j}}{||\boldsymbol{a_i}||_2 \cdot ||\boldsymbol{a_j}||_2} \tag{14}$$

In (14), $a_i$ denotes the i$^{\text{th}}$ column of the matrix $\boldsymbol{A}$ i.e., a spatial impulse response, and each column was normalized by its L2-norm to omit the influence of the amplitude variations. For a clear perception, **Fig. 6** (b) shows the cosine similarity matrix of the spatial impulse responses shown in **Fig. 6** (a). This means that the diagonal elements in **Fig. 6** (b) are unity i.e., 1 by definition, representing the auto-correlation coefficient of each response, while the off-diagonal elements denote the pairwise cross-correlation coefficients between different responses, with values in the range of [-1, 1]. Thus, each row or column of matrix $\boldsymbol{C}$ quantifies the spatial coherence of one voxel among all the others in the lateral plane introduced by the coded aperture.

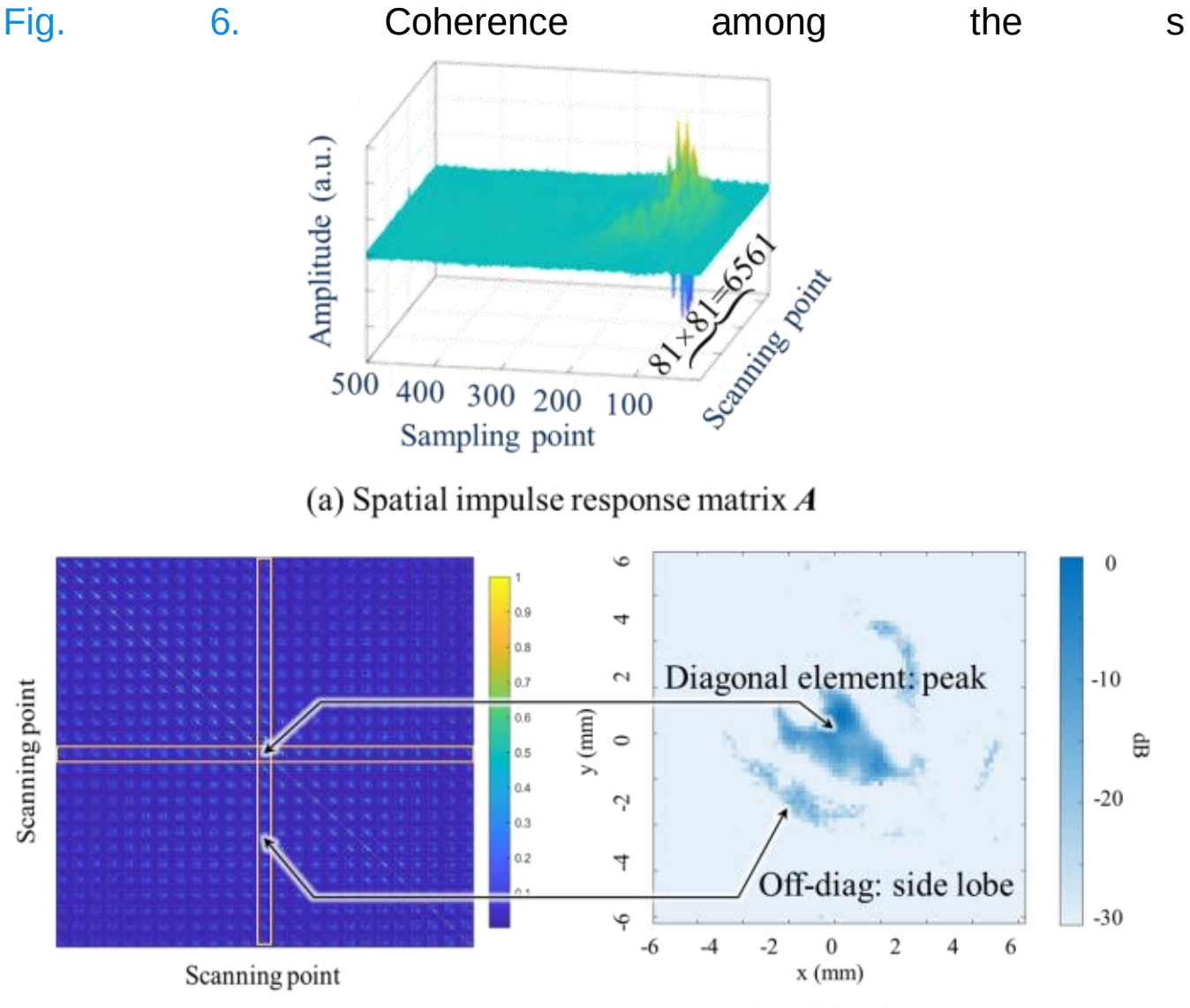

Fig. 6. Coherence among the spatial impulse responses. (a) An exemplary set of spatial impulse response functions generated by a coded aperture at a given depth. (b) Cosine similarity matrix of the signals in (a). (c) Schematic demonstration of the spatial coherence.

To visualize and quantify such spatial coherence among the response functions, each row of matrix $\boldsymbol{C}$ was reshaped back into a 2D spatial map according to the voxel coordinates, as illustrated in **Fig. 6** (c). The -6 dB area of the main lobe was then calculated to define the spatial coherence area, over which coded echo signatures remain highly correlated and are therefore difficult to be distinguished. It is however noted that the spatial coherence area differs fundamentally from the actual imaging resolution cell. The former is the intrinsic property of a coding mask under a single reference voxel condition, while the latter is affected by inter-voxel interference arising from the simultaneous presence of multiple scatterers or objects in practical imaging scenarios. Qualitatively, a smaller spatial coherence area indicates that different voxels possess more distinguishable echo signatures, indicating the superior encoding performance of a mask, which is desirable for compressive imaging.

To link the spatial coherence and mask design, **Fig. 7** provides a statistic of the spatial coherence of the masks over depth ($z$), in form of box plot. For a direct comparison with the pixel side length, we show here the equivalent spatial coherence length of each mask, defined as the square root of the coherence area.

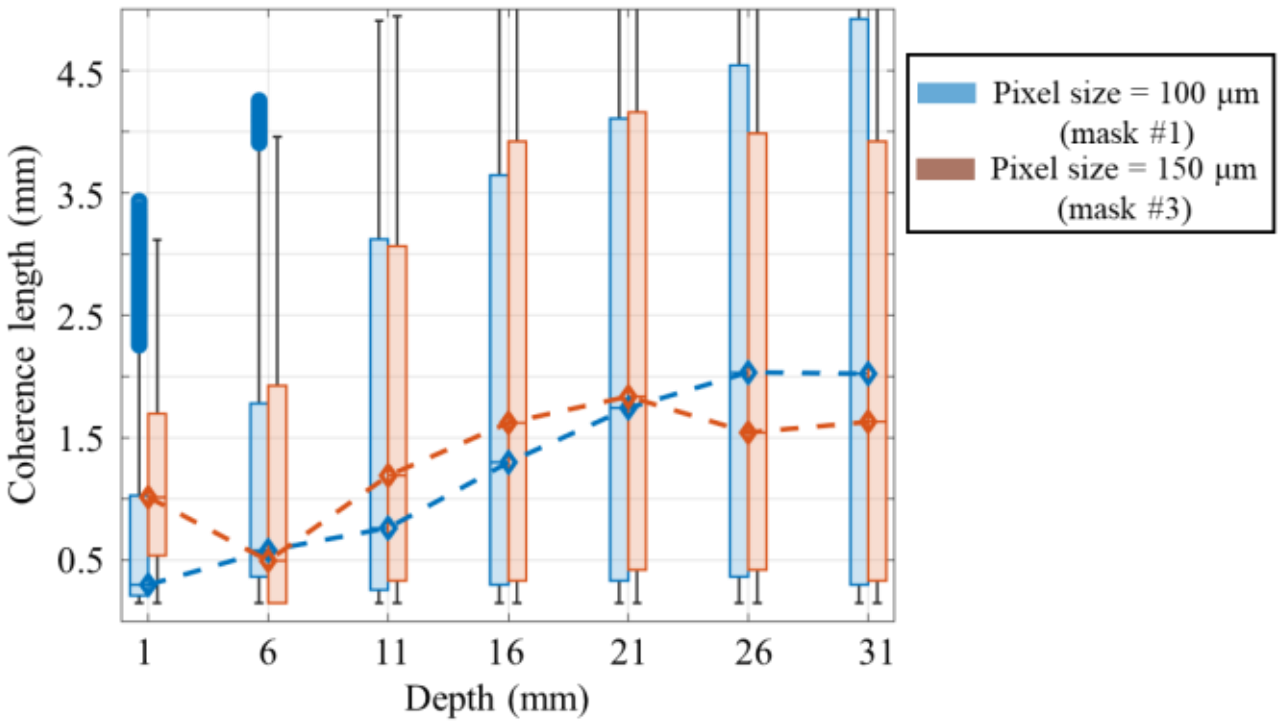

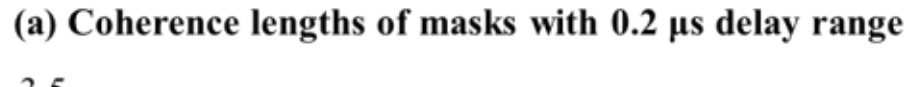
(a) Coherence lengths of masks with 0.2 μs delay range

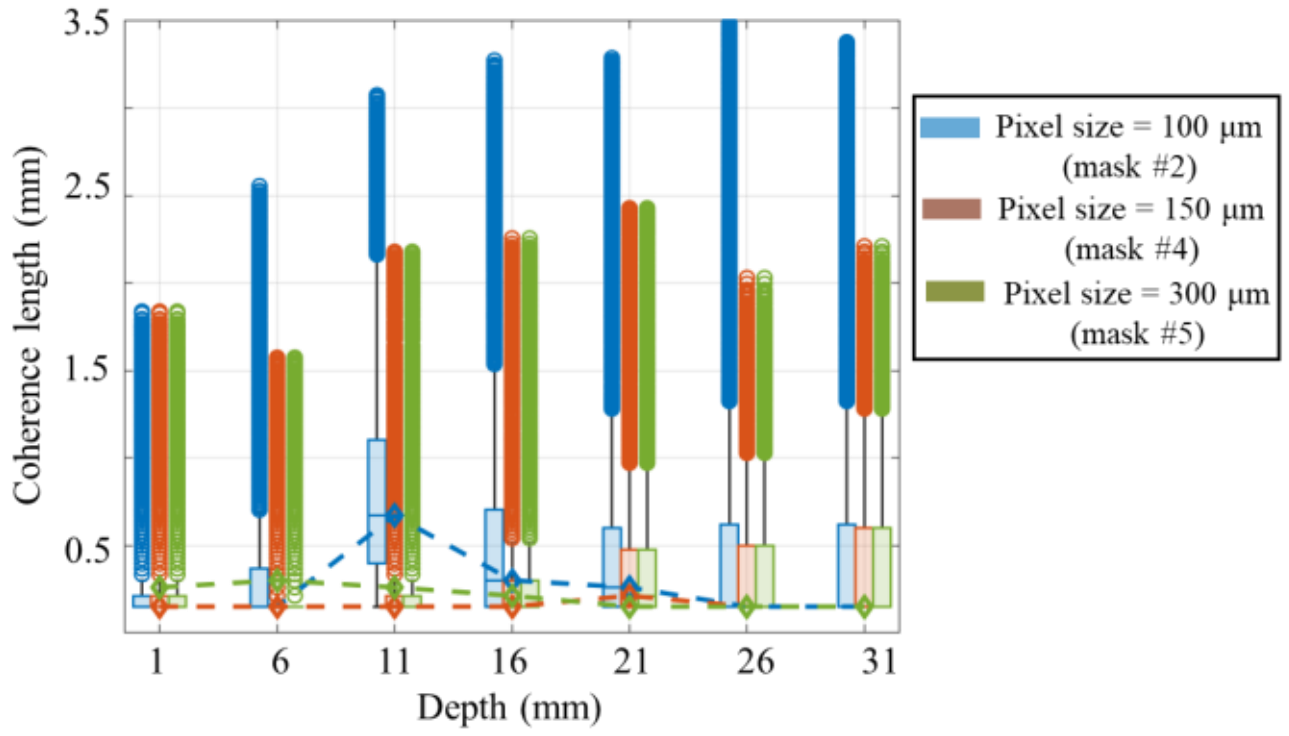

(b) Coherence lengths of masks with 0.4 μs delay range

Fig. 7. Spatial coherence lengths of the different masks: (a) with time-delay range of 0.2 μs; and (b) of 0.4 μs.

Two trends related to mask design can be observed. First, for the masks with the smaller time delay range of 0.2 μs (masks #1 and #3), the equivalent coherence length gradually increases with depth, indicating that the coded echo signatures lose their diversity as the wavefront propagates. In contrast, the other three masks with the larger time delay range of 0.4 μs (masks #2, #4 and #5) exhibit a more stabilized coherence length over the imaging depth within the entire near field of the transducer. In addition, despite the presence of outliers, the median coherence lengths of masks #2, #4 and #5 remain comparable to the corresponding mask pixel side lengths, as marked by the dashed lines. This result suggests that the scale of spatial coherence is primarily determined by the pixel size, while larger delay range helps to stabilize this coherence scale over depth.

### 2) *Information capacity of the different masks*

From an information theory perspective, the actual information capacity of each mask is further quantified using the entropy-based effective rank of the cosine similarity matrix ***C*** [38]. As this provides an SBP-like measure of the information capacity of a coded aperture, it enables a direct comparison with the desired SBP of an imaging system. Specifically, the independent spatial modes that are encoded in the cosine similarity matrix is characterized by its eigenvalue spectrum. The eigenvalues of matrix ***C*** are first normalized to form a probability distribution over the eigenmodes [38], as given in (15) where $\lambda_k$ denotes the k-th eigenvalue of matrix ***C***, and $n$ is the number of spatial sampling points in the calibrated plane.

$$p_k = \frac{\lambda_k}{\sum_{j=1}^{j=n} \lambda_j} \tag{15}$$

The spectral entropy is then calculated as (16) [38].

$$H(C) = -\sum_{k=1}^{k=n} p_k \log(p_k) \tag{16}$$

The entropy-based rank of matrix ***C*** characterizing the effective number of independent spatial modes is defined as (17) [38].

$$M_{eff} = e^{H(C)} \tag{17}$$

This definition indicates that the information capacity increases when the eigenvalue spectrum is distributed more uniformly over a larger number of modes. Finally, to compare the information capacities of masks with different dimensions under a unified framework, we define a dimensionless coding capacity factor as (18), in which $n$ is the number of spatial points on a calibrated plane.

$$\delta = M_{eff}/n \tag{18}$$

Following this characterization pipeline, **Fig. 8** depicts the information capacities of the different masks over calibrated depth. It can be observed that the masks with the smaller time-delay range of 0.2 μs generally provide lower and more depth-dependent encoding capacity. This indicates that insufficient delay diversity cannot maintain sufficiently independent coded echo signatures as the wavefront propagates. In contrast, the masks with the larger time-delay range of 0.4 μs provide higher and more stable encoding capacity over depth, suggesting that stronger mask-induced wavefield modulation enhances the global independence of the spatial impulse responses.

In addition, for masks with the same delay range, smaller pixel sizes tend to provide higher encoding capacity, which is consistent with the reduced spatial coherence length discussed above. This result suggests that the time-delay range mainly affects the depth stability of encoding, whereas the pixel size controls the spatial scale of independent encoding modes.

Overall, the five masks investigated here provide $\delta$ values in the range of 0.15 % – 1.1 %, indicating that only a small fraction of the sampled spatial degrees of freedom can be independently encoded in a single-shot measurement.

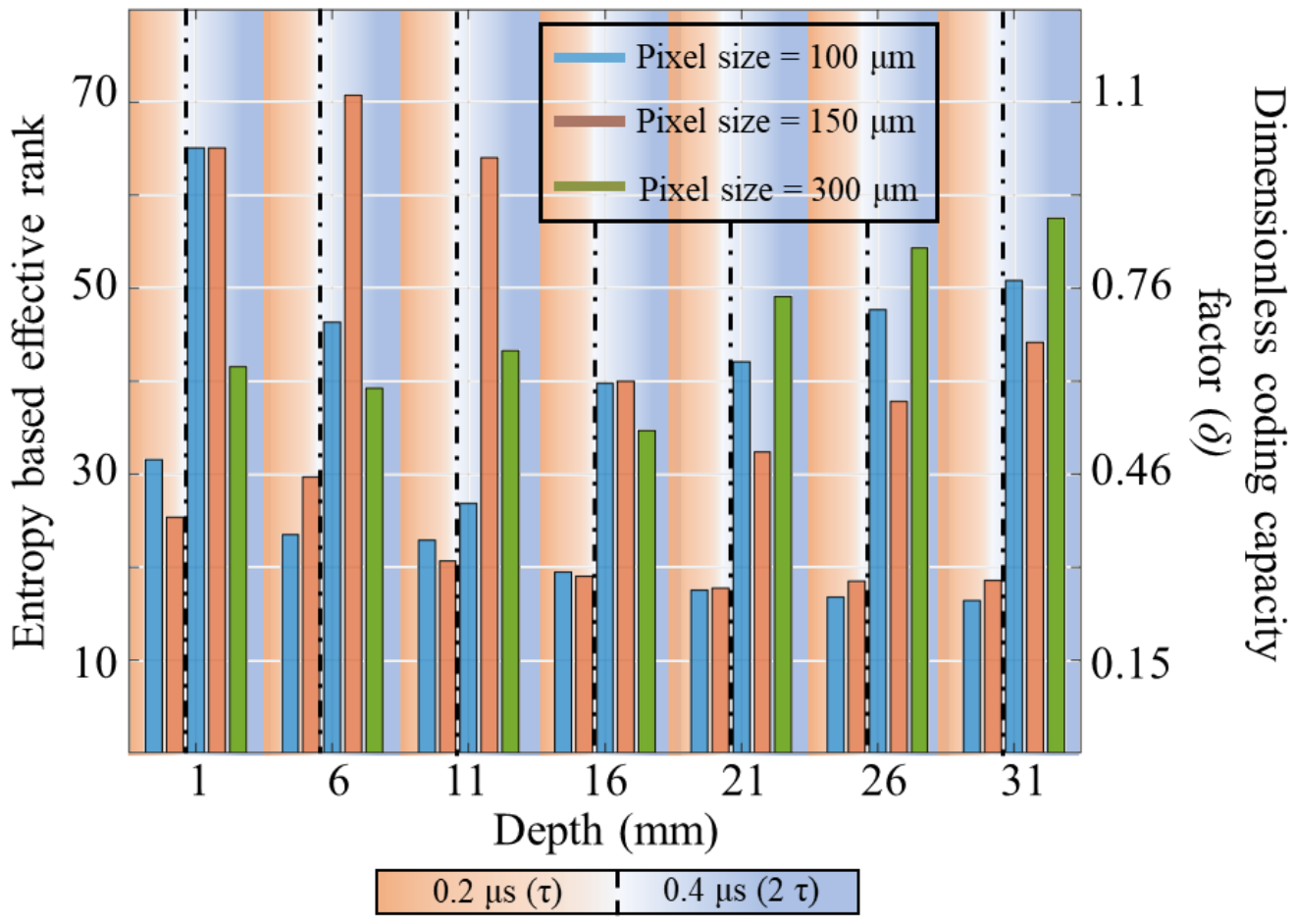

Fig. 8. The information capacities of the different masks over imaging depth that are characterized by the entropy based effective rank and dimensionless coding capacity factor.

Since the encoding capacity varies not only among masks but also over imaging depth, each mask-depth combination is treated as an individual encoding system in the following recoverability analysis.

## C. Performance of particle localization towards ULM

Ultrasound localization microscopy (ULM) resolves fine structures and fluid dynamics by tracking localized tracer particles over frames. Thus, the ULM imaging speed highly relies on the concentration of tracer particles that can be resolved in each acquisition, and the reconstruction time of each snapshot. For compressive imaging, a higher concentration of tracer particles reduces the required accumulation time, but arises the inter-particle interference in the encoded RF signals. Aiming for assessing the ability of compressive imaging using random masks toward high-speed ULM, it is crucial to determine the number of particles that can be resolved simultaneously under a given encoding capacity. In particular, we investigated the reconstruction performance of matched filter and sparsity constrained reconstruction using FISTA under a same encoding capacity.

Since different masks provide different encoding capacities over imaging depths, we introduce another dimensionless number $\rho$, which represents the fraction of the available encoding capacity occupied by simultaneously present particles as given in (19). This dimensionless parameter enables a unified comparison of reconstruction performance across different masks.

$$\rho = \frac{k}{M_{eff}} \tag{19}$$

In (19), $k$ is the number of tracer particles in each accumulated frame.

We evaluated the reconstruction performance of $\rho$ values in a range of 0.01 – 0.6, using a Monte-Carlo sampling method. Specifically, for each $\rho$ value, 200 random test images were generated by placing one-hot point scatterers at randomly selected voxel locations. The voxel size was half wavelength, i.e., 150 μm at 5 MHz in water, so that the echo from a sub-wavelength particle can be eligibly represented by the spatial impulse response from a point scatterer in the corresponding voxel. For each test image, the single-snapshot RF signal was synthesized following the process given in (1).

**Fig. 9** shows an example of the ground-truth particle distribution corresponding to ($\delta$, $\rho$) of (0.011, 0.285), the synthetic RF signal at different SNRs, and the corresponding reconstructions using the matched filter and FISTA solved sparsity-constrained LS problem, respectively. It is observed that the matched filter [**Fig. 9** (c) – (e)] not only reconstructed many false-positive particles that did not present in the ground-truth [**Fig. 9** (a)], but also missed many particles that were actually present. This is because the non-optimized random masks naturally give rise to inter-voxel coherence via their spatial impulse response functions, and the Gram operation employed by the matched filter is not able to decouple such coherence within the spatial coherence length. Furthermore, the matched filter suffers from spatial amplitude bias where regions with stronger response amplitudes dominate the reconstruction. Therefore, the particles located in weak response regions may be suppressed. Nevertheless, the SNR possessed a negligible influence on the matched filter reconstructions.

In contrast, the sparsity-constrained LS reconstruction allows the encoded temporal signatures to be separated more effectively and accurately from the acquired signals. As shown in **Fig. 9** (f) & (g), at moderate and high SNRs i.e., 40 dB and 60 dB, all the 20 particles in the ground-truth were correctly reconstructed, despite minor residual artefacts. However, this method is more sensitive to noise. As the SNR decreases to 20 dB, the artefacts became unacceptable in the raw reconstruction [**Fig. 9** (h)], which visually hinders particle identification and localization.

Aiming for a systematic evaluation of the reconstruction performance of the two methods, we further calculated the percentage of the correctly localized particles in each test image for all the ($\delta$, $\rho$) pairs addressed in this study. For each reconstruction, the same number of particles as in the ground-truth was selected from the reconstructed image by taking the $K$ peaks with the highest intensities, where $K$ is same the number of particles in ground-truth. The percentage of the correctly localized particles in each reconstruction is therefore defined as (20), in which $N_{correct}$ is the number of correctly localized particles in the reconstruction.

$$\sigma = \frac{N_{correct}}{K} \tag{20}$$

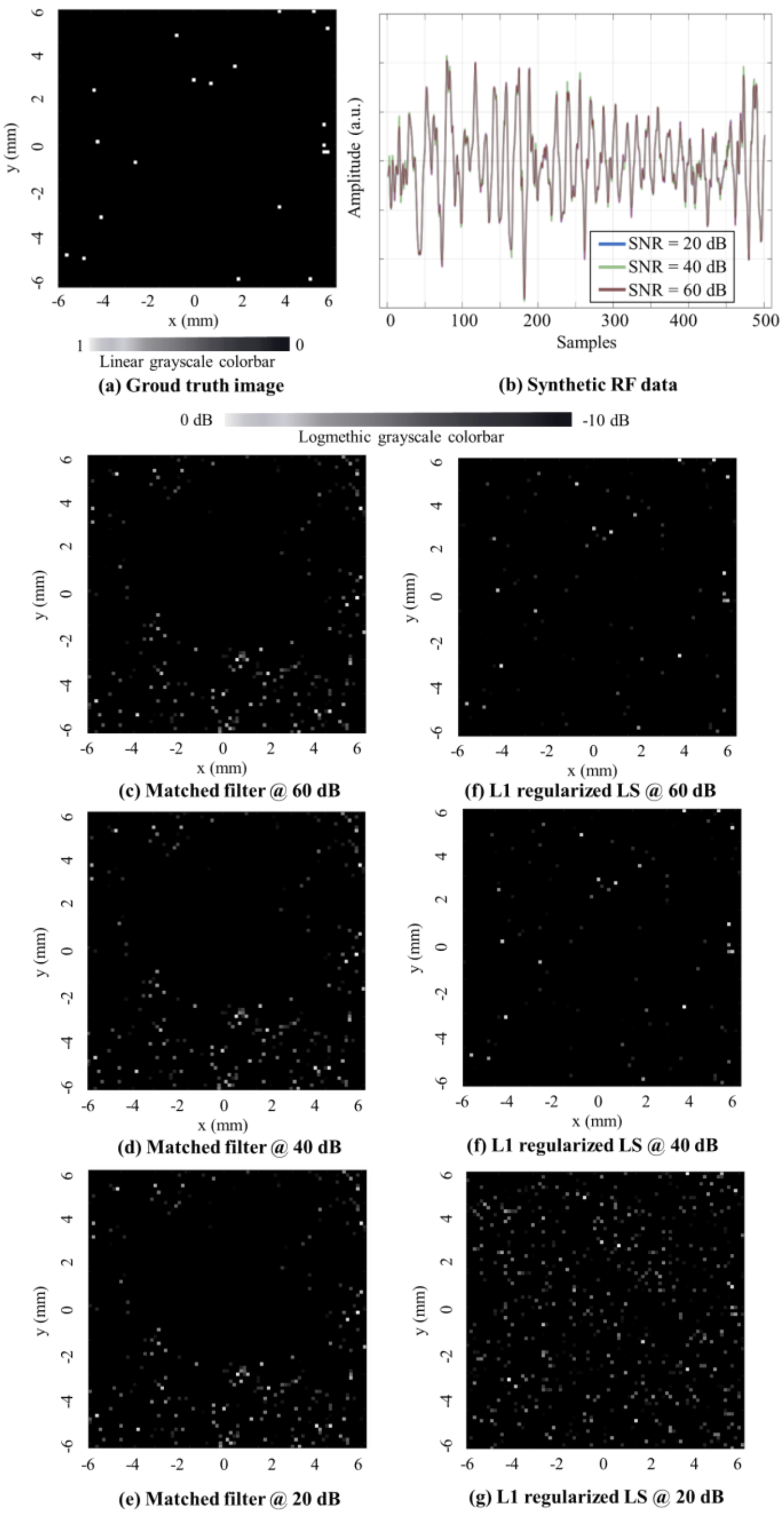

Fig. 9. Demonstrations of matched filter and sparsity constrained LS reconstructions for particle localization at different SNRs. (a) Ground-truth containing 20 particles. (2) Synthetic RF data at different SNRs of 20, 40 and 60 dB. (c) - (e) Matched filter reconstructions at the three SNRs. (f) – (h) Sparsity-constrained LS reconstructions at the three SNRs.

The average value of the percentage of the correctly localized particles over the 200 Monte-Carlo samplings indicates the probability of successful particle localization, and therefore provides a measure for evaluating the particle localization accuracy. **Fig. 10** plots this localization accuracy of the two reconstruction methods for each ($\delta$, $\rho$) pair at a SNR of 40 dB, in which each pixel in the map characterizes the probability of successful particle localization under a given encoding capacity and particle concentration.

**Fig. 10** (a) shows that, for matched filtering, the localization accuracy remains low over most of the investigated ($\delta$, $\rho$) pairs. Only a small number of particles, corresponding to approx. 1% of the dimensionless coding capacity provided by a mask, can be correctly localized under relatively favorable conditions, as indicated by the red arrows in **Fig. 10** (a). In other words, under the encoding capacity of the current random aberration masks, matched filter is only suitable for single- or very few-particle localization regimes, which drastically limits its practical applications in practical ULM. This result also aligns with the fact that matched filtering is strongly limited by Gram operation induced inter-voxel interference and spatial amplitude bias. Given these, although the computational efficiency of matched filter enables millisecond level single snapshot reconstruction, it is not suitable yet for localizing multiple particles using the current non-optimized random coding masks.

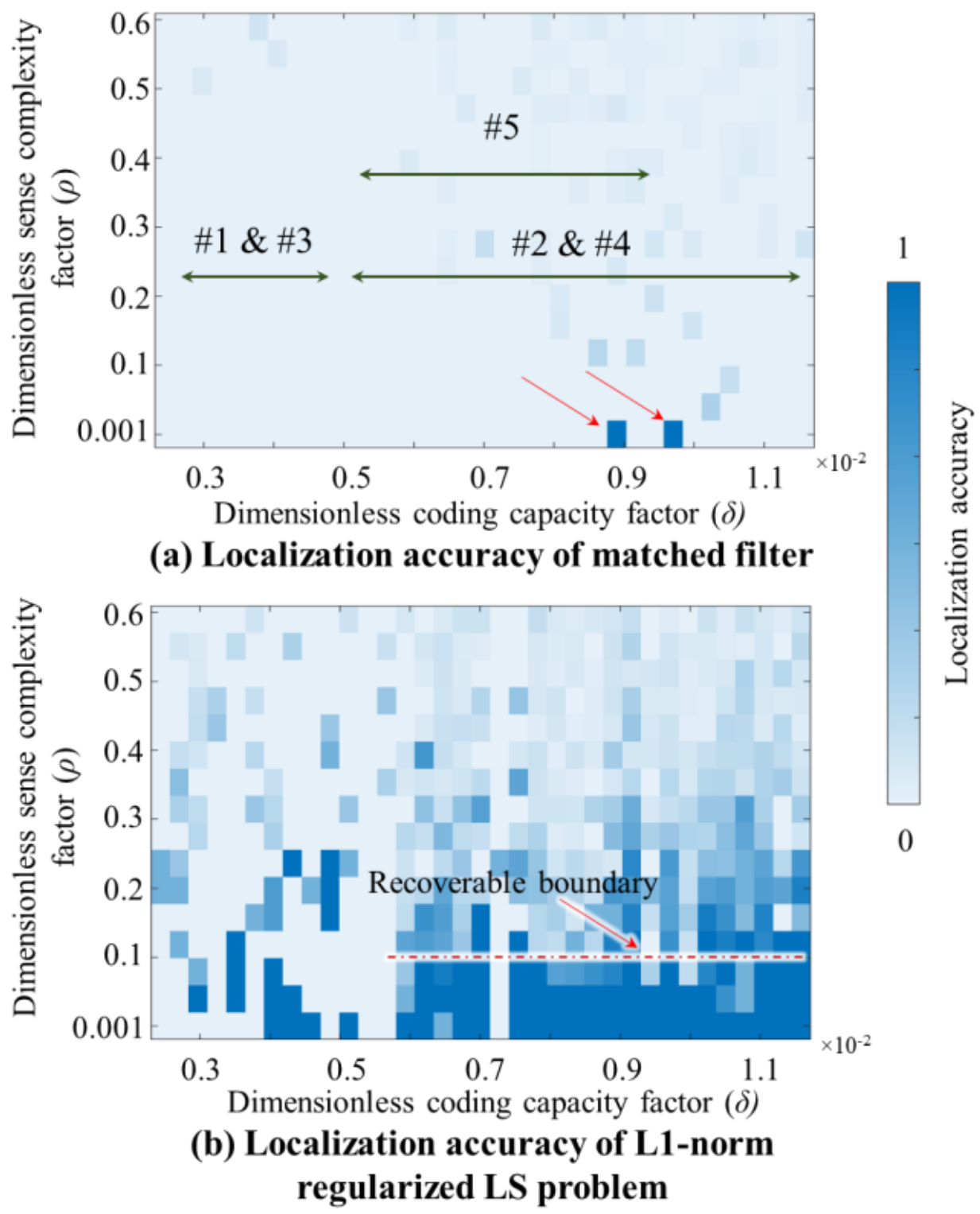

Fig. 10. Localization accuracy of sub-wavelength particles at an SNR of 40 dB: (a) using the matched filter; (b) using FISTA solved L1-norm regularized LS problem.

Compared with matched filtering, the sparsity-constrained LS reconstruction provides a much broader recoverable region in the ($\delta$, $\rho$) space. As shown in **Fig. 10** (b), the number of particles corresponding to around 10% of the encoding capacity i.e., $\delta$ can be faithfully reconstructed in a single snapshot. As $\rho$ further increases, more particles simultaneously present in the scene, leading to stronger inter-particle interference in the single-snapshot RF measurement, decreasing the localization accuracy. As a result, a transition between the successful and failed reconstruction zones manifests in **Fig. 10** (b). In other words, the encoding capacity of the mask determines the system-side information budget, while the reconstruction algorithm determines how efficiently this budget can be utilized.

Nevertheless, the time complexity of sparsity-constrained LS reconstruction remains a critical challenge. The reconstruction time of a single snapshot is on the order of hundreds of milliseconds in the current implementation, which is approximately two orders of magnitude slower than matched filtering.

These findings provide practical guidelines towards compressive imaging based ULM. First, although matched filtering is attractive for real-time ULM, mask optimization, towards lower coherence and more uniform amplitude distribution among the spatial impulse response functions, is necessary before it can be used as a practical high throughput particle localization method. Second, for sparsity-constrained LS reconstruction, the proposed ($\delta$, $\rho$) representation provides a quantitative criterion for selecting the tracer particle concentration. Given the recoverable boundary of approx. 10 % of the encoding capacity provided by a mask, once the mask is characterized, the particle concentration can be selected within the recoverable regime, to make full use of the available encoded information, meanwhile omitting unnecessarily high inter-voxel interferences. Third, expanding the information capacity of the mask is always beneficial for increasing the absolute number of particles that can be simultaneously localized, which improves the temporal resolution of compressive ULM.

### D. Performance of B-mode imaging

To further evaluate the performance of B-mode imaging, a 3D digital ultrasound phantom was generated on the same spatial grid as the calibrated forward model. The phantom consisted of 81×81×233 voxels and contained a speckle-like background together with representative structural features, including hyperechoic (high-contrast) inclusions and hypoechoic (low-echo) regions. These features were designed to mimic typical contrast variations in B-mode ultrasound images [39]. **Fig. 11** (a) shows representative slices of the 3D digital phantom, which ensures a non-sparse structural imaging task that is distinct from the isolated point-scatterer case used for particle localization.

**Fig. 11** (b) shows exemplary reconstructions of synthetic data from mask #5 at 21 mm, using matched filtering, LSQR, and TV-regularized LS reconstruction. Due to the limited information capacity of the current non-optimized coding masks, all three methods resulted in degraded and visually ambiguous reconstructions for such complex structures. In this particular example shown in **Fig. 11** (b), it is observed that the matched filter failed completely, which is consistent with its strong Gram operation induced inter-voxel interference artefacts. The TV-regularized LS reconstruction was only able to resolve the hypoechoic regions with the sharp boundaries, but not the hyperechoic regions and the speckle background. In contrast, LSQR was able to resolve both the hypoechoic and hyperechoic regions, except the speckle background, which outperforms the other two methods.

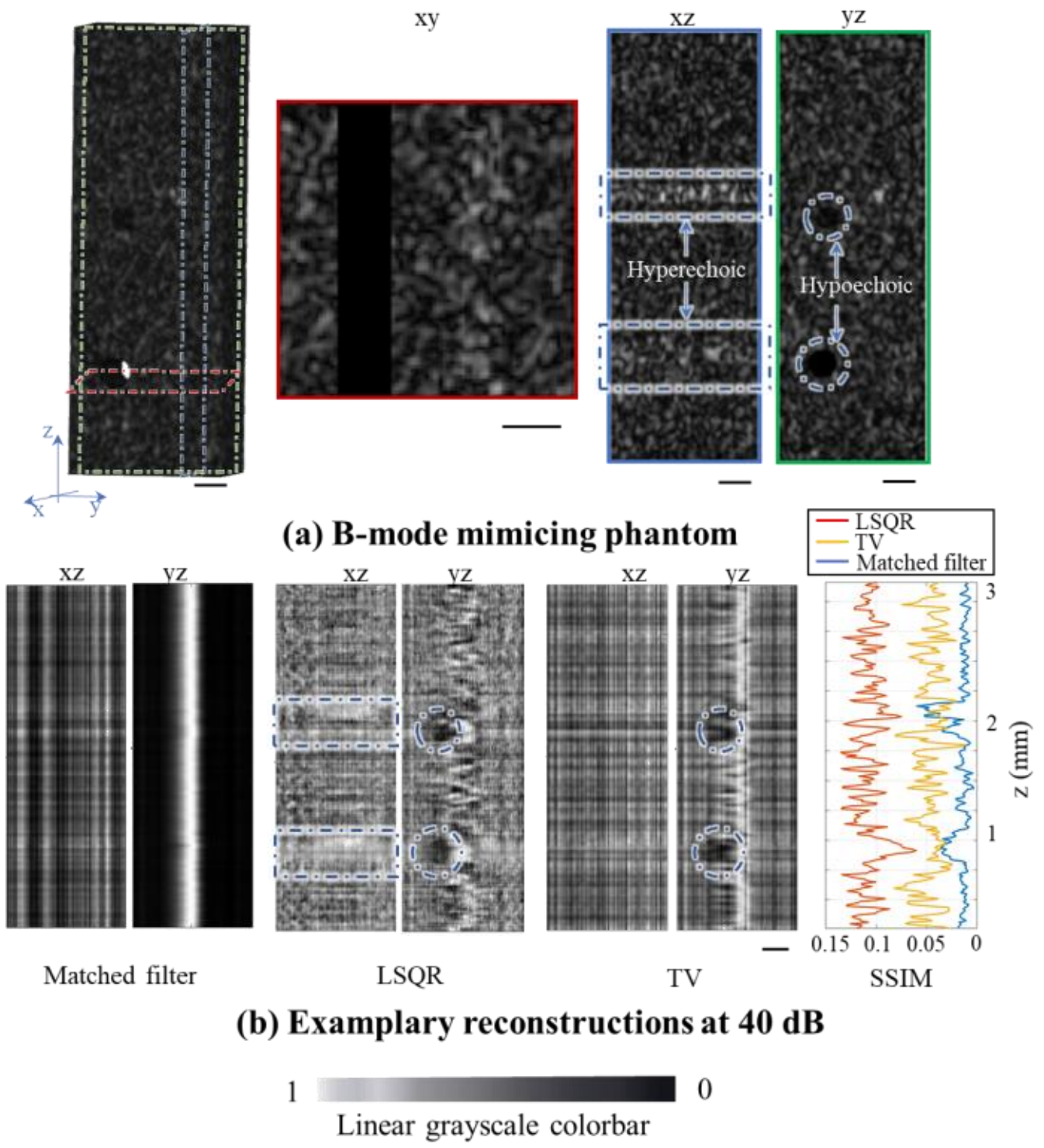

Fig. 11. Demonstrations of compressive imaging of a B-mode mimicking phantom (Scale bar of 2 mm.). (a) The B-mode mimicking phantom, in which the different views, the hyperechoic and hypoechoic regions are indicated. (b) Representative reconstructions of the different methods at an SNR of 40 dB, in which their SSIM values over depth are also plotted.

In order to further quantify the reconstruction quality, the structural similarity index measure (SSIM) was calculated between each reconstructed image and the ground-truth phantom, as given in (21) [40].

$$\mathrm{SSIM}(x,\hat{x}) = \frac{(2\mu_x \cdot \mu_{\hat{x}} + c1)\cdot(2\sigma_{x\hat{x}} + c2)}{({\mu_x}^2 + {\mu_{\hat{x}}}^2 + c1)\cdot({\sigma_x}^2 + {\sigma_{\hat{x}}}^2 + c2)} \quad (21)$$

In (21), $x$ and $\hat{x}$ denote the ground-truth and reconstructed images, respectively; $\mu_x$ and $\mu_{\hat{x}}$ are their mean intensities, ${\sigma_x}^2$ and ${\sigma_{\hat{x}}}^2$ are their variances, and $\sigma_{x\hat{x}}$ is their covariance; *c1* and *c2* are small positive constants used to ensure numerical stability when the denominator is close to zero. The value of SSIM ranges between 0 and 1, and the higher the value the more similar between the reconstruction and ground-truth. The SSIM values of the reconstructed *xy* slices over depth ($z$) using the three methods w.r.t the ground-truth are also plotted in **Fig. 11** (b), in which the matched filter exhibits the lowest SSIM, LSQR provides the highest SSIM, while intermediate SSIM is realized by TV-regularized LS reconstruction. This is also consistent with the visual observations in the reconstructions shown in **Fig. 11** (b).

Finally, to obtain the relation between encoding capacity and structuring imaging performance, **Fig. 12** plots volumetric SSIM of B-mode imaging as a function of the dimensionless coding capacity factor $\delta$. For each reconstruction method, the shaded band represents the SSIM range over the tested SNR levels, while the dashed line shows the representative result at 40 dB. Overall, LSQR provides the highest SSIM over most of the investigated configurations, followed by TV-regularized LS reconstruction and matched filter. Furthermore, as $\delta$ increases, the SSIM upper envelopes of LSQR and TV-regularized LS reconstruction are improved, indicating a positive correlation between encoding capacity and structural reconstruction quality. However, the SSIM does not increase monotonically with $\delta$, especially in the intermediate-capacity region. This is because such a complicated phantom structure results in an ill-posed reconstruction problem that exceeds the information capacity provided by a mask, where many voxels are simultaneously activated, and the reconstruction quality is therefore affected by the detailed inter-voxel interference of the full forward operator.

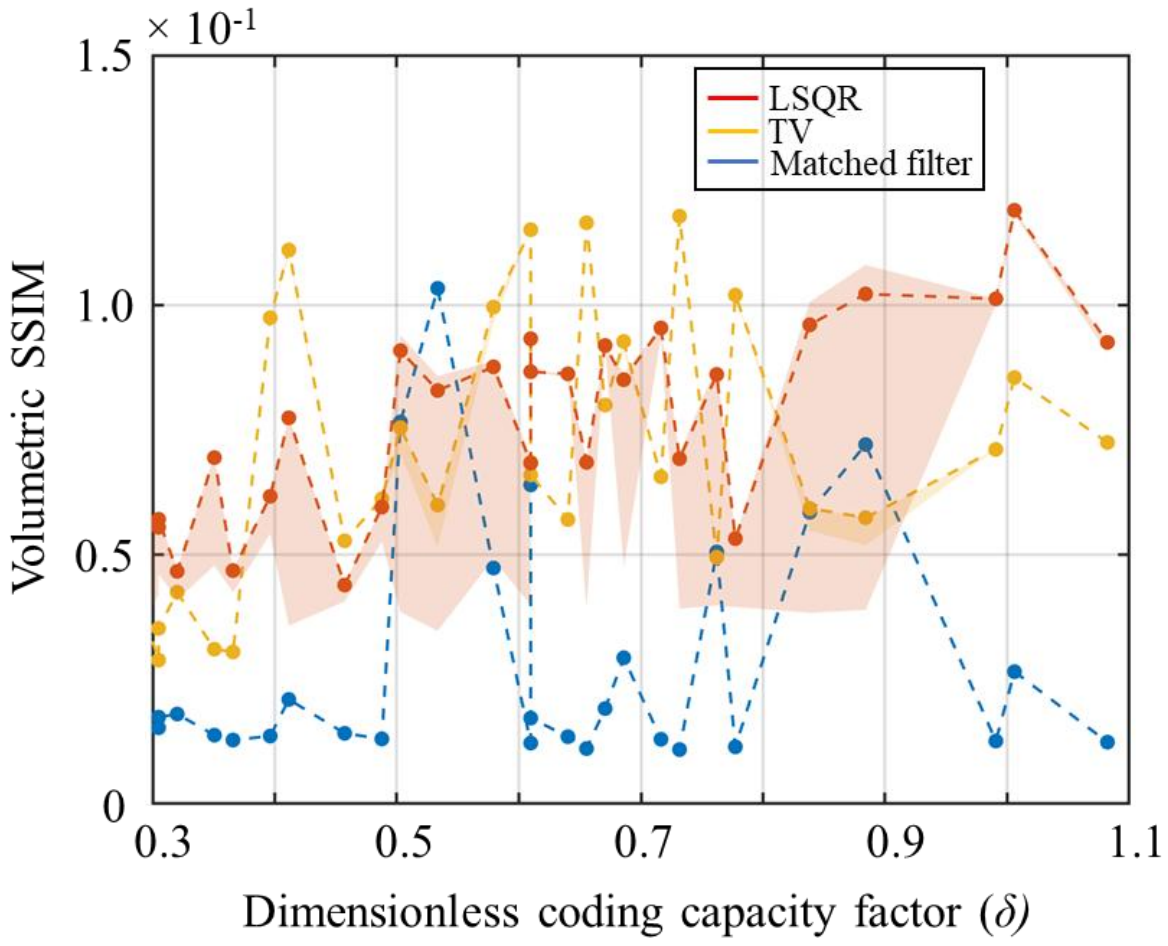

Fig. 12. The relation among volumetric SSIM, information capacity and SNR of the different reconstruction methods for structural compressive imaging.

Meanwhile, the width of the shaded bands reflects the SNR sensitivity of the different reconstruction methods. Matched filtering remains insensitive to SNR (no shaded band present in **Fig. 12**), but its SSIM stays low because of the incapability of the Gram operation for decoupling the inter-voxel coherence. TV-regularized LS reconstruction provides intermediate performance, with moderate improvement over matched filtering and limited SNR dependence. In contrast, LSQR exhibits the largest variation across SNR levels: when the SNR is sufficiently high e.g., over 30 dB, LSQR can exploit more of the encoded spatial information and achieves higher SSIM; while at lower SNRs e.g., below 20 dB, noise amplification reduces its advantage.

Despite the superior performance of LSQR and TV-regularized LS reconstructions, their reconstruction qualities using the current non-optimized random coding masks remain insufficient for practical structural imaging. Furthermore, both reconstruction methods can be implemented at 10's ms level for one 2D slice. Therefore, the temporal resolution of 3D imaging in a volume containing e.g., 100 slices would be at second level, which is slower than desired by ultrafast imaging. These deficiencies are mainly attributed to the limited information capacity encoded by the masks that is far below what is required for reliable recovery of complex structures. This leads to the inapplicability of the computationally efficient matched filter, and long iterating process using LSQR and TV-regularized LS reconstructions. This limitation motivates mask optimization towards higher information capacity for future single-snapshot compressive ultrasound imaging.

### *E. Summary and discussions*

A central outcome of this work is the $(\delta, \rho)$ representation for evaluating the SBP of compressive imaging systems, which decouples the intrinsic properties of the encoding and decoding paths. Specifically, the coded aperture defines the available information budget, while the reconstruction algorithm determines how efficiently this budget can be extracted i.e., the fraction of this budget occupied by the object to be reconstructed. This allows the performance of different compressive imaging systems to be compared in a unified framework, without relying on individual mask design or imaging phantom.

The mask characterization further reveals how random mask design parameters influence the encoding capacity. In general, smaller pixel sizes e.g., of half wavelength combined with larger time-delay range e.g., of two periods at the carrier frequency are beneficial to reduce the spatial coherence, thereby providing higher information capacity. However, the current non-optimized random masks still provide insufficient information capacity for high-quality single-snapshot 4D imaging. Therefore, further mask optimization should focus on increasing the number of spatial encoding modes, and on enabling a more uniform sensitivity distribution over the imaging region. In other words, an optimized mask must improve the orthogonality among the spatial impulse response functions of a mask, so that the coded spatial information becomes more accessible to computationally efficient reconstruction methods e.g., the matched filter.

For this purpose, a quantitative model linking mask design parameters to the information capacity is necessary. As the present results show that the pixel size and time-delay range provide two important design parameters, but they do not fully determine the actual encoding capacity. The final capacity depends on the complete wavefield formation process, including diffraction and RF bandwidth. Therefore, future mask design should be guided by the measured or simulated forward operator rather than by the geometrical parameters alone.

## V. Conclusions and Outlook

Single-snapshot compressive ultrasound imaging provides a promising paradigm toward high temporal resolution volumetric imaging with substantially reduced hardware complexity. The practical performance of compressive imaging is fundamentally governed by the information capacity encoded by a mask and how efficiently the information can be restored by the reconstruction algorithm.

In order to evaluate the performance of single-snapshot compressive ultrasound imaging systems under a unified framework, we introduced two dimensionless parameters, namely the dimensionless coding capacity factor and the dimensionless scene complexity factor.

The main contribution of this work is therefore threefold. First, it provides a physical interpretation of compressive ultrasound imaging performance in practical imaging scenarios. Second, it enables a practical criterion for selecting operating parameters, such as the suitable tracer particle concentration in ULM relevant particle localization. However, for dense structural imaging, the reconstruction quality is also affected by the full inter-voxel interference structure of the forward operator. Third, it provides design guidance for future coding masks by separating the intrinsic distinguishability of encoded temporal signatures from amplitude-weighted sensitivity effects.

Looking ahead, future high-quality ultrafast single-snapshot compressive ultrasound imaging, toward higher coding capacity and higher scene complexity, should be understood and designed as a joint encoding–reconstruction system. One route is to optimize the mask for the computationally efficient matched filter reconstruction, as its millisecond-level reconstruction time is attractive for 4D imaging. In this scene, the Welch inequality provides a useful theoretical basis of the lower bound of mutual coherence [41], but practical mask optimization must additionally account for bandwidth, diffraction, and uniformity of pulse-echo sensitivity.

A second route is to preserve the LS based reconstruction methods while reducing their computational cost. Deep unfolding or learned iterative reconstruction may provide a promising direction by converting iteration processes into fast neural-network realizations, while retaining interpretability from the underlying inverse problem [42], [43], [44].

Finally, practical deployments of compressive ultrasound imaging also require more robust and scalable calibration strategies. The current approach relies on experimentally measured spatial impulse responses, which can be time-consuming and sensitive to mask fabrication tolerances and mounting conditions. Future work should therefore investigate calibration-free or calibration-light implementations, including electronic or reconfigurable coding schemes that can generate controllable spatial encoding patterns without relying solely on passive 3D-printed masks. Such developments would further improve the generalization, adaptability, and practical applicability of single-snapshot compressive ultrasound imaging in a wide range of applications.

## Acknowledgment

The authors would like to thank David Weik, Hannes Bischoff, Konrad Liess, and Yuezhen Xu, for valuable discussions. The authors greatly thank Andreas Frölich from Horizon Microtechnologies GmbH, for providing necessary experimental materials.

## Data Availability

The authors declare that all data supporting the findings of the study are presented within this article.

## References

[1] J. Provost *et al.*, "3D ultrafast ultrasound imaging in vivo," *Phys. Med. Biol.*, vol. 59, no. 19, pp. L1–L13, 2014, doi: 10.1088/0031-9155/59/19/L1.

[2] G. Park *et al.*, "Fetal monitoring for high-risk pregnancies using a wearable ultrasound patch," *Nat. Biotechnol.*, pp. 22–25, 2026, doi: 10.1038/s41587-026-03140-1.

[3] C. Wang and X. Zhao, "See how your body works in real time — wearable ultrasound is on its way," *Nature*, vol. 630, no. 8018, pp. 817–819, 2024, doi: 10.1038/d41586-024-02066-5.

[4] D. Weik *et al.*, "Current Trends in Ultrasound Wearables : Spotlight on System Architecture," *IEEE Rev. Biomed. Eng.*, vol. PP, pp. 1–21, 2026, doi: 10.1109/RBME.2026.3664011.

[5] B. G. Sgambato *et al.*, "Virtual reality interactions via a user-generic ultrasound human-machine interface for wrist and hand tracking," pp. 1–16, 2025.

[6] Z. Dou, Y. Xu, Z. Wei, H. Emmerich, J. Czarske, and D. Weik, "Towards Monitoring Water Content in Membrane Electrode Assembly of Low Temperature Fuel Cells Using Ultrasound," in *2024 IEEE Ultrasonics, Ferroelectrics, and Frequency Control Joint Symposium (UFFC-JS)*, 2024, pp. 1–4. doi: 10.1109/UFFC-JS60046.2024.10794159.

[7] Z. Dou *et al.*, "A Water Monitoring System for Proton Exchange Membrane Fuel Cells Based on Ultrasonic Lamb Waves: An Ex Situ Proof of Concept," *IEEE Trans. Instrum. Meas.*, vol. 72, pp. 1–12, 2023, doi: 10.1109/TIM.2023.3329101.

[8] Z. Dou, L. Tropf, H. Hoster, H. Schmidt, J. Czarske, and D. Weik, "Advanced Ultrasonic Diagnostic Technology Towards Green Hydrogen Energy Systems," in *IEEE International Ultrasonics Symposium, IUS*, 2023, pp. 1–4. doi: 10.1109/IUS51837.2023.10307141.

[9] Z. Dou *et al.*, "Multiscale fluid transport in low-temperature electrolyzers and fuel cells : Linking transport phenomena with analytical technologies towards improved performance," *Chem. Eng. J.*, vol. 544, p. 178919, 2026, doi: 10.1016/j.cej.2026.178919.

[10] L. Gruter, R. Nauber, and J. W. Czarske, "Ultrasonic Bubble Imaging in Molten Salt Using a Multi-Mode Waveguide and Time Reversal," *IEEE Trans. Instrum. Meas.*, vol. 71, 2022, doi: 10.1109/TIM.2022.3144739.

[11] C. Othmani, K. Ließ, D. Räbiger, S. Eckert, J. Czarske, and L. Büttner, "Mode Analysis in Ultrasound Multimode Waveguides Toward Computational Imaging," *IEEE Trans. Instrum. Meas.*, vol. 75, 2026, doi: 10.1109/TIM.2026.3690805.

[12] Y. Tang, R. Tsumura, J. T. Kaminski, and H. K. Zhang, “Actuated Reflector-Based 3-D Ultrasound,” *IEEE Trans. Ultrason. Ferroelectr. Freq. Control*, vol. 69, no. 8, pp. 2437–2446, 2022, doi: 10.1109/TUFFC.2022.3180980.

[13] Z. Dou *et al.*, “Scanning Acoustic Microscopy for Quantifying Two-phase Transfer in Operando Alkaline Water Electrolyzer,” *J. Power Sources*, vol. 660, p. 238575, 2025, doi: 10.1016/j.jpowsour.2025.238575.

[14] Y. Lyu, Y. Shen, M. Zhang, and J. Wang, “Real-Time 3D Ultrasound Imaging System Based on a Hybrid Reconstruction Algorithm,” *Chinese J. Electron.*, vol. 33, no. 1, pp. 245–255, 2024, doi: 10.23919/cje.2023.00.002.I.

[15] M. Caudoux *et al.*, “3D ultrafast imaging using a 3072-element matrix array,” *IEEE Trans. Ultrason.*, vol. PP, p. 1, 2026, doi: 10.1109/TUSON.2026.3680661.

[16] C. Rabut *et al.*, “4D functional ultrasound imaging of whole-brain activity in rodents,” *Nat. Methods*, vol. 16, no. October, 2019, doi: 10.1038/s41592-019-0572-y.

[17] G. Matrone *et al.*, “A Volumetric CMUT-Based Ultrasound Imaging System Simulator With Integrated Electronics Models,” *IEEE Trans. Ultrason. Ferroelectr. Freq. Control*, vol. 61, no. 5, pp. 792–804, 2014, doi: 10.1109/TUFFC.2014.2971.

[18] C. Risser *et al.*, “Real-time volumetric ultrasound research platform with 1024 parallel transmit and receive channels,” *Appl. Sci.*, vol. 11, no. 13, 2021, doi: 10.3390/app11135795.

[19] A. Ramalli, E. Boni, E. Roux, H. Liebgott, and P. Tortoli, “Design, Implementation, and Medical Applications of 2-D Ultrasound Sparse Arrays,” *IEEE Trans. Ultrason. Ferroelectr. Freq. Control*, vol. 69, no. 10, pp. 2739–2755, 2022, doi: 10.1109/TUFFC.2022.3162419.

[20] J. A. Jensen *et al.*, “Anatomic and Functional Imaging Using Row – Column Arrays,” *IEEE Trans. Ultrason. Ferroelectr. Freq. Control*, vol. 69, no. 10, pp. 2722–2738, 2022, doi: 10.1109/TUFFC.2022.3191391.

[21] N. Haidour *et al.*, “Multi-lens ultrasound arrays enable large scale three-dimensional micro- vascularization characterization over whole organs,” *Nat. Commun.*, vol. 16, p. 9317, 2025, doi: 10.1038/s41467-025-64911-z.

[22] P. Kruizinga *et al.*, “Compressive 3D ultrasound imaging using a single sensor,” *Sci. Adv.*, vol. 3, no. 12, 2017, doi: 10.1126/sciadv.1701423.

[23] J. Janjic *et al.*, “Structured ultrasound microscopy,” *Appl. Phys. Lett.*, vol. 112, no. 25, 2018, doi: 10.1063/1.5026863.

[24] C. Othmani *et al.*, “Waveguide-based in-process ultrasound inspection and imaging in harsh environments : advances and challenges,” *Ultrasonics*, vol. 168, p. 108214, 2026, doi: 10.1016/j.ultras.2026.108214.

[25] Y. Hu *et al.*, “A 240 Elements Matrix Probe with Aberration Mask for 4D Carotid Artery Computational Ultrasound Imaging,” *IEEE Trans. Ultrason. Ferroelectr. Freq. Control*, 2025, doi: 10.1109/TUSON.2025.3648199.

[26] M. D. Brown *et al.*, “Four-dimensional computational ultrasound imaging of brain hemodynamics,” *Sci. Adv.*, vol. 10, no. 3, pp. 1–11, 2024, doi: 10.1126/sciadv.adk7957.

[27] Y. Hu, M. Brown, E. Mc, E. Mc, and J. G. Bosch, “Compressive Imaging with Spatial Coding Masks on Low Number of Elements : An Emulation Study,” in *2022 IEEE International Ultrasonics Symposium (IUS)*, pp. 1–4. doi: 10.1109/IUS54386.2022.9957699.

[28] G. A. Guarneri *et al.*, “A Sparse Reconstruction Algorithm for Ultrasonic Images in Nondestructive Testing,” *Sensors*, vol. 15, pp. 9324–9343, 2015, doi: 10.3390/s150409324.

[29] S. Gao *et al.*, “A TV regularisation sparse light field reconstruction model based on guided-filtering, ” *Signal Processing: Image Communication*, vol. 109, 2022, doi: 10.1016/j.image.2022.116852.

[30] K. Melde, A. G. Mark, T. Qiu, and P. Fischer, “Holograms for acoustics,” *Nature*, vol. 537, no. 7621, pp. 518–522, 2016, doi: 10.1038/nature19755.

[31] Z. Dou, *Adaptive Ultrasound Imaging of Fluid Transport in Green Hydrogen Energy Systems*. Düren: Shaker Verlag GmbH, 2025. [Online]. Available: https://www.shaker.de/de/site/content/shop/index.asp?lang=de&ID=8&ISBN=978-3-8191-0426-8

[32] Z. Dou, D. Karnaushenko, O. G. Schmidt, and D. Karnaushenko, “A High Spatiotemporal Resolution Ultrasonic Ranging Technique with Multiplexing Capability,” *IEEE Trans. Instrum. Meas.*, vol. 70, 2021, doi: 10.1109/TIM.2021.3120129.

[33] A. Beck and M. Teboulle, “A Fast Iterative Shrinkage-Thresholding Algorithm,” *SIAM J. IMAGING Sci.*, vol. 2, no. 1, pp. 183–202, 2009, doi: 10.1137/080716542.

[34] C. C. Paige and M. A. Saunders, “LSQR : An Algorithm for Sparse Linear Equations and Sparse Least Squares,” *ACM Trans. Math. Softw*, vol. 8, no. 1, pp. 43–71, 1982.

[35] “lsqr Solve system of linear equations — least-squares method.” https://de.mathworks.com/help/matlab/ref/lsqr.html

[36] H. Tao and Z. Li, “An alternative minimization method for TV-image deblurring in tensor space,” *Appl. Math. Model.*, vol. 145, no. August 2024, p. 116102, 2025, doi: 10.1016/j.apm.2025.116102.

[37] J. Duran, B. Coll, and C. Sbert, “Chambolle ’ s Projection Algorithm for Total Variation Denoising,” vol. 3, pp. 311–331, 2013.

[38] O. Roy and M. Vetterli, “THE EFFECTIVE RANK : A MEASURE OF EFFECTIVE DIMENSIONALITY,” no. Eusipco, pp. 606–610, 2007.

[39] B. Shin, S. Jeon, J. Ryu, and H. J. Kwon, “Compressed Sensing for Elastography in Portable Ultrasound,” *Ultrason. Imaging*, pp. 1–22, 2017, doi: 10.1177/0161734617716938.

[40] I. Bakurov, M. Buzzelli, R. Schettini, M. Castelli, and L. Vanneschi, “Structural similarity index ( SSIM ) revisited : A data-driven approach,” *Expert Syst. Appl.*, vol. 189, no. August 2020, p. 116087, 2022, doi:

10.1016/j.eswa.2021.116087.

[41] L. R. Welch, “Lower Bounds on the Maximum Cross Correlation of Signals,” *IEEE Trans. Inf. Theory*, vol. 20, no. 3, pp. 397–399, 1973, doi: 10.1109/TIT.1974.1055219.

[42] T. Glosemeyer, Y. Ma, R. Kuschmierz, J. Wu, L. Cao, and J. W. Czarske, “Real-time physics-informed neural network image reconstruction for a see-through camera via an AR lightguide,” *Adv. Imaging*, vol. 2, no. 6, 2025, doi: 10.3788/AI.2025.10000.

[43] F. Wang, J. W. Czarske, and G. Situ, “Deep learning for computational imaging: from data-driven to physics-enhanced approaches,” *Adv. Photonics*, vol. 7, no. 5, 2025, doi: 10.1117/1.AP.7.5.054002.

[44] Y. Gao and L. Cao, “Model - based deep learning enables time - resolved computational microscopy,” *PhotonicX*, vol. 7, no. 3, 2026, doi: 10.1186/s43074-025-00222-2 PhotoniX.